\documentclass[structabstract]{aa} 
\usepackage{graphicx}
\usepackage{txfonts}
\usepackage[utf8]{inputenc}
\usepackage{dcolumn}
\usepackage{supertabular}
\usepackage{rotating}
\usepackage{longtable}
\usepackage{longtable, lscape}
\usepackage{lscape}
\usepackage{float}
\usepackage{subfig}
\usepackage{url}
\usepackage{color}
\usepackage{natbib}
\usepackage[dvipsnames]{xcolor}
\usepackage[normalem]{ulem}
\usepackage{soul}

\newcommand{\kms}{\mbox{km s$^{-1}$}}
\newcommand{\hco}{\mbox{H$^{13}$CO$^{+}$}}

\newcolumntype{d}[1]{D{.}{\cdot}{#1}}
\newcolumntype{.}{D{.}{.}{-1}}
\newcommand{\hh}{H$_{2}$}

\newcommand{\hii}{H{\sc ii}}
\newcommand{\hiis}{H{\sc ii}s}
\newcommand{\uchii}{UC\,H{\sc ii}}

\newcommand{\KS}{Kolmogorov-Smirnov}
\newcommand{\mum}{$\mu$m}

\newcommand{\msun}{M$_\odot$}

\newcommand{\co}{C$^{18}$O}
\newcommand{\cch}{C$_{2}$H}
\newcommand{\hcni}{H$^{13}$CN}
\newcommand{\hcnii}{HC$^{15}$N}
\newcommand{\hnc}{HN$^{13}$C}
\newcommand{\hccch}{c-C$_{3}$H$_{2}$}

\begin{document} 
\title{ATLASGAL-selected massive clumps in the inner Galaxy:} 
\subtitle{VIII. Chemistry of photodissociation regions}
\authorrunning{Kim et al.}
\titlerunning{PDRs in dust clumps}


   \author{W.-J. Kim
          \inst{1, 2}
          \and
		  F. Wyrowski\inst{1}
          \and
          J. S. Urquhart\inst{3}
          \and 
          J. P. Pérez-Beaupuits\inst{4}
          \and
          T. Pillai\inst{5,1}
          \and
          M. Tiwari\inst{6,1}
          \and
          K. M. Menten\inst{1}
          }
            
  \institute{ Max-Planck-Institut f\"ur Radioastronomie, Auf dem H\"ugel 69, 53121 Bonn, Germany
  \and Present address: Instituto de Radioastronom\'ia Milim\'etrica, Avenida Divina Pastora 7, 18012 Granada, Spain\\ \email{kim@iram.es} 
  \and  School of Physical Sciences, University of Kent, Ingram Building, Canterbury, Kent CT2\,7NH, UK
  \and  European Southern Observatory, Alonso de C\'ordova 3107, Vitacura Casilla 7630355, Santiago, Chile
  \and  Institute for Astrophysical Research, 725 Commonwealth Ave, Boston University Boston, MA 02215, USA
  \and  Department of Astronomy, University of Maryland, College Park, MD 20742, USA} 
\date{Received 29 July 2020 / Accepted 28 September 2020 }

 
  \abstract
   {}
   {We study ten molecular transitions obtained from an unbiased 3\,mm molecular line survey using the IRAM 30\,m telescope toward 409 compact dust clumps identified by the APEX Telescope Large Area Survey of the Galaxy (ATLASGAL) to understand photodissociation regions (PDRs) associated with the clumps. The main goal of this study is to investigate whether the  abundances of the selected molecules show any variations resulting from the PDR chemistry in different clump environments.}
   {We selected HCO, HOC$^+$, \cch, \hccch, CN, \hcni, \hcnii, and \hnc\ as PDR tracers, and \hco\ and \co\ as dense gas tracers. By using estimated optical depths of \cch\ and \hcni\ and assuming optically thin emission for other molecular transitions, we derived column densities of those  molecules and their abundances. To assess the influence of the presence and strength of ultra-violet radiation, we compare abundances of three groups of the clumps: \hii\ regions, infrared bright non-\hii\ regions, and infrared dark non-\hii\ regions.}
   {We detected \co, \hco, \cch, \hccch, CN and \hnc\ toward most of the observed dust clumps (detection rate $>$ 94\%), and \hcni\ is also detected with a detection rate of 75\%. On the other hand, HCO and \hcnii\ show detection rates of 32\% and 39\%, respectively, toward the clumps, which are mostly associated with \hii\ region sources: detection rates of HCO and \hcnii\ toward the \hii\ regions are 66\% and 79\%. We find that the abundances of HCO, CN, \cch, and \hccch\ decrease as the \hh\ column density increase, indicating high visual extinction, while those of high density tracers (i.e., \hco\ and \hcnii) are constant. In addition, $N$(HCO)/$N$(\hco) ratios significantly decrease as \hh\ column density increase, and in particular, 82 clumps have $X$(HCO) $\gtrsim 10^{-10}$ and $N$(HCO)/$N$(\hco) $\gtrsim 1$, which are the indication of far-ultraviolet (FUV) chemistry. This suggests the observed HCO abundances are likely associated with FUV radiation illuminating the PDRs. We also find that high $N$(\hccch)/$N$(\cch) ratios found for \hii\ regions having high HCO abundances ($\gtrsim\,10^{-10}$) are associated with more evolved clumps with high $L_{\rm bol}$/$M_{\rm clump}$. This trend might be associated with gain-surface processes, which determine initial abundances of these molecules, and time-dependent effects in the clumps corresponding to the envelopes around dense PDRs and \hii\ regions. In addition, some fraction of the measured abundances of the small hydrocarbons of the \hii\ sources can be the result of the photodissociation of PAH molecules.}
   {}


\maketitle

\section{Introduction}

Newly born high-mass stars ($>$\,8\,\msun) dramatically affect their environment in various ways, e.g. by powerful outflows, creation of ionized gas regions via ultraviolet (UV) radiation, and stellar winds, etc. \citep{zinnecker2007}. Their intense UV radiation impinging on the surrounding molecular gas, creates photodissociation regions (PDRs, e.g., \citealt{tielens1985,tielens2013}). The UV inducing photo-chemistry and many other factors, such as the geometric structure of molecular gas, volume density of \hh, and turbulence, control the formation and destruction of molecules leading to a complex chemistry in PDRs. So far, most detailed studies of PDRs have concentrated on a few nearby regions (e.g., the Horsehead nebula and the Orion Bar; \citealt{rodriguez-franco1998,teyssier2004,pety2005,gerin2009,cuadrado2015}).

According to the detailed molecular excitation studies of \cite{hogerheijde1995} and \cite{jansen1995} of the Orion Bar, the PDR layer consists of at least two components: a low-density inter-clump medium ($n$(\hh) $\approx 3\times10^{4}$\,cm$^{-3}$) and high-density clumps ($n$(\hh) $\approx 1\times 10^{6}$\,cm$^{-3}$). More recent observational studies have indicated the presence of dense PDRs between the ionized and neutral gas in the Orion Bar and Monoceros R2 (Mon R2) \citep{rodriguez-franco1998,rizzo2005}. Especially toward Mon R2, high \hh\ densities ($>4\times10^{6}$\,cm$^{-3}$) were found in its PDRs \citep{rizzo2005}, which are illuminated by intense UV radiation ($G_{0}\footnote{The UV radiation field is in Habing untits, which are a measure of the  average far-UV interstellar radiation field; $G_0 =1$ corresponds to a flux of $1.6\times10^{-3}$~erg~cm$^{-2}$~s$^{-1}$.} = 5\times10^{5}$ ; \citealt{rizzo2003}) unlike low- and moderate-UV illuminated PDRs (e.g., the Horsehead nebula) ($G_{0} = 100 - 2600$; \citealt{pilleri2013}), and they are geometrically thin ($\lesssim 0.001$\,pc). In the other extreme, PDRs surrounding compact \hii\ or ultracompact (\uchii) regions in molecular clumps have different physical conditions than the aforementioned ``weak'' PDRs, such as that associated with the Horsehead nebula. Their high densities suggest that \uchii\ regions could be pressure-confined by the surrounding high-density neutral gas \citep{rizzo2005}. The observations toward Mon R2 provided a good understanding of a particular type of dense PDR illuminated by high-UV radiation ($h\nu > $13.6\,eV) from embedded high-mass stars. It is, however, hard to build a comprehensive view of such PDRs by studying only one object. Therefore, investigating a robust sample is important that is not biased by any evolutionary stages, and with access to various molecular species to identify appropriate tracers of different layers of the PDRs.

In our previous studies \citep{kim2017, kim2018}, we identified \hii\ regions toward compact dust clumps selected from the ATLASGAL compact source catalogues \citep{contreras2013,urquhart2014_atlas_cata} using hydrogen radio recombination lines (RRLs) at (sub)millimeter wavelengths and radio continuum surveys. The clumps with RRL detections are associated with more energetic UV radiation fields from mostly O-stars, whereas the clumps with only radio continuum associations may have weaker UV fields provided by early B-stars. Many of the remaining clumps without those signposts have embedded massive young stellar objects (MYSOs) that provide even weaker radiation fields. Finally, some of the clumps without embedded objects, which are classified as infrared (IR) dark clumps, may only be affected by cosmic-ray in their interior or by external radiation fields from nearby star-forming complexes. The ATLASGAL compact source catalogues, therefore, allow us to investigate PDRs produced in a wide range of UV field strengths and physical conditions.

In general, dense and compact PDRs occur in complex star-forming regions that are embedded in their parental molecular clouds. It is critical to find molecular tracers whose abundance is primarily driven by PDR chemistry. Infrared diagnostics provide direct information about PDRs, but they are not easily observable from the ground and accessible in the presence of high extinction. Alternatively, several molecular species are linked to UV photochemistry without suffering high extinction and also offer additional information on the velocity fields that allow us to investigate the kinematics and turbulence in PDRs. \cite{rizzo2005} identified two groups of molecules as PDR tracers: the first group is related to the surface layers of the PDR (visual extinction 2\,mag\,$<A_{V}<$\,5\,mag) that are exposed to a high UV field (referred to as high UV or HUV). Molecular species in this group are reactive ions (e.g., CO$^{+}$ and HOC$^{+}$) and small hydrocarbons (e.g., \cch\ and \hccch; \citealt{sternberg1995}). The second group is found in deeper parts of PDRs (5\,mag\,$< A_{V} <$\,10\,mag) where they are less exposed to the UV field (i.e., low UV, 6\,eV $< h\nu < $13.6\,eV; referred to as FUV) and close to cold gas regions \citep{schilke2001,rizzo2005}. Radicals (e.g., HCO, \cch\ and CN) are considered to belong to this group. 

In this paper, we analyze 3\,mm line survey data taken with the Institut de Radioastronomie Millim\'etrique (IRAM) 30\,m telescope toward 409 ATLASGAL dust clumps. These observations contain a large number of rotational transitions of molecular species. Based on previous observational studies (e.g., \citealt{rizzo2005, boger2005, gerin2009}) we selected eight molecules that are regarded as typical PDR tracers: HCO, HOC$^{+}$, \cch, \hccch, \hcni, \hcnii, \hnc\ and CN. In addition, \co\ and \hco\ were chosen as general probes of column density and dense gas in the dust clumps. The selected molecular transitions are given in Table\,\ref{tb:mol_info}.

\begin{table*}
\centering
\small
\caption{\label{tb:mol_info}The observed molecular transitions.}
\begin{tabular}{ l l c c c c c c c }
\hline \hline
Species & Transition & Frequency & $g_{\rm u}$& $E\rm_{up}$ & $\mu$ &$S\rm_{ij}^{\dagger}$ & Relative $I^{b}$ & $A\rm_{ij}$ \\
 &     & [MHz] & & [K] & [Debye] & & &[s$^{-1}$] \\
\hline
\co\  & $J=1-0$ & 109782.173   & 3.0 & 5.3 & 0.11 & 1.000 & & $6.27\times10^{-8}$\\\cline{1-9}
HCO &$N_{\rm K_{a}, K_{c}} = 1_{0, 1}-0_{0, 0}, J=3/2-1/2, F= 2-1$ & 86670.760 & 5.0 & 4.2 &1.36 & 1.666 & 0.421 & $4.69\times10^{-6}$ \\
HCO &$N_{\rm K_{a}, K_{c}} =1_{0, 1}-0_{0, 0}, J=3/2-1/2, F= 1-0$ & 86708.360 & 3.0 & 4.2 & & 0.979 & 0.247 & $4.60\times10^{-6}$ \\
HCO &$N_{\rm K_{a}, K_{c}} =1_{0, 1}-0_{0, 0}, J=1/2-1/2, F= 1-1$ & 86777.460 & 3.0 & 4.2 & & 0.979 & 0.247 & $4.61\times10^{-6}$ \\   
HCO &$N_{\rm K_{a}, K_{c}} =1_{0, 1}-0_{0, 0}, J=1/2-1/2, F= 0-1$ & 86805.780 & 1.0 & 4.2& & 0.333 & 0.084 & $4.71\times10^{-6}$ \\ 
\hline
\hco\ & $J=1-0$ &  86754.288&3.0&  4.2 & 3.90 &1.000 & & $3.85\times10^{-5}$ \\\cline{1 -9}
HOC$^{+}$ & $J=1-0$ & 89487.414 &3.0& 4.3 & 2.77 & 1.000 & & $7.20\times10^{-5}$ \\ 
\hline
\cch\ &  $N=1-0, J=3/2-1/2, F=1-1$ & 87284.156 &3.0& 4.2&0.77 & 0.170 & 0.043 & $2.81\times10^{-7}$  \\ 
\cch\ &  $N=1-0, J=3/2-1/2, F=2-1$ & 87316.925 &5.0& 4.2& & 1.667 & 0.417 & $1.65\times10^{-6}$  \\ 
\cch\ &  $N=1-0, J=3/2-1/2, F=1-0$ & 87328.624 &3.0& 4.2 && 0.830 & 0.208 & $1.37\times10^{-6}$  \\ 
\cch\ &  $N=1-0, J=1/2-1/2, F=1-1$ & 87402.004 &3.0& 4.2 && 0.830 & 0.208 & $1.38\times10^{-6}$  \\ 
\cch\ &  $N=1-0, J=1/2-1/2, F=0-1$ & 87407.165 &1.0& 4.2 && 0.333 & 0.083 & $1.66\times10^{-6}$  \\ 
\cch\ &  $N=1-0, J=1/2-1/2, F=1-0$ & 87446.512 &3.0& 4.2 && 0.170 & 0.043 & $2.82\times10^{-7}$  \\\cline {1 -9}
\hccch\ & $J_{\rm K_{a}, K_{c}}=2_{1,2}-1_{0,1}$& 85338.894 &15.0& 6.4 & 3.27 & 4.503 &  & $2.32\times10^{-5}$ \\ 
\hline
\hcni\ & $J=1-0, F=1-1$ &  86338.767 & 3.0 &  4.1& 2.99 &1.067 & 0.333  & $2.22\times10^{-5}$ \\
\hcni\ & $J=1-0, F=2-1$ &  86340.184 & 5.0 &  4.1&&1.778 & 0.555 & $2.22\times10^{-5}$ \\
\hcni\ & $J=1-0, F=0-1$ &  86342.274 & 1.0 &  4.1&&0.356 & 0.111 & $2.22\times10^{-5}$ \\\cline{1-9}
\hcnii\ &  $J=1-0$ &  86054.966 & 3.0 & 4.1 & 2.99 & 1.000 & & $2.20\times10^{-5}$\\\cline{1-9}
\hnc\  & $J=1-0$ & 87090.850 & 3.0 & 4.2 & 2.70 &1.000 & & $1.87\times10^{-5}$ \\\cline{1-9}
CN & $N=1-0, J=1/2-1/2, F=1/2-1/2$  &  113123.370&2.0&   5.4&1.45& 0.073 &0.037& $1.29\times10^{-6}$ \\
CN & $N=1-0, J=1/2-1/2, F=1/2-3/2$  &  113144.157&2.0&   5.4&& 0.594 &0.297& $1.05\times10^{-5}$ \\
CN & $N=1-0, J=1/2-1/2, F=3/2-1/2$  &  113170.492&4.0&   5.4&& 0.580 &0.290& $5.15\times10^{-6}$ \\
CN & $N=1-0, J=1/2-1/2, F=3/2-3/2$  &  113191.279&4.0&   5.4&& 0.753 &0.377& $6.68\times10^{-6}$ \\
\hline
\end{tabular}
\tablefoot{($\dagger$) $S\rm_{ij}$ is the line strength taken from the JPL and CDMS catalogs.
(b) Expected relative intensities ($S_{ij}$/$\sum\,S_{ij}$), assuming that lines are optically thin ($\tau_{\nu} < 1$). The sum of relative intensities of the hyperfine lines is normalized to 1. $g_{u}$ is the statistical weight of the upper state level, and $E_{\rm u}$ is the energy of the upper level of the selected transition. $\mu$ is the permanent dipole moment of the species. $A_{ul}$ is the Einstein coefficient for spontaneous emission.}
\end{table*}

The structure of the paper is as follows. The selected molecules are described in Section \ref{ch:lines}. The observations, source types, and data reduction are explained in Section \ref{ch:obs_data}. Detected molecular lines and detection rates are presented in Section \ref{ch:results}, including a description of several sources with CN self-absorption line profiles. The estimated column densities and abundances of the selected molecules are given in Section \ref{ch:analysis} along with a comparison with \hh\ column densities. Column density ratios of some of the selected molecular lines (i.e., HCO, \hco, \cch, and \hccch) and their correlations are discussed in Section \ref{ch:pdr_mol}. Finally, we summarize our main results in Section \ref{ch:summary}.

\begin{table*}
\begin{center}
\small
\caption[]{\label{tb:sou} List of observed sources. }
\begin{tabular}{l c c c c c c c}
\hline\hline
ATLASGAL  & RA. & Dec. & Dist. & $T_{\rm dust}$ & Type & Classification$^{\dagger}$& Comments \\
 name & $\alpha$(J2000) & $\delta$(J2000) & (kpc) & (K) &  &  & \\
\hline
AGAL006.216$-$00.609 & 18:02:02.9 & $-$23:53:13 &  3.0 & 14.6 & non-\hii\ &        24 dark &   \\
AGAL008.049$-$00.242 & 18:04:35.2 & $-$22:06:40 & 10.9 & 18.2 & non-\hii\ & IR bright or \hii\ &   \\
AGAL008.671$-$00.356 & 18:06:19.0 & $-$21:37:28 &  4.4 & 25.9 &     \hii\ & IR bright or \hii\ &  no \co, CN, \hcni, \hcnii\ data  \\
AGAL008.684$-$00.367 & 18:06:23.0 & $-$21:37:11 &  4.4 & 24.5 & non-\hii\ &        24 dark & no \cch\ data  \\
AGAL008.706$-$00.414 & 18:06:36.6 & $-$21:37:16 &  4.4 & 11.9 & non-\hii\ & IR bright or \hii\ &   \\
\hline
\end{tabular}
\tablefoot{($\dagger$) The temperature of dust, distance, and classification are taken from \cite{urquhart2018}. This table is a fraction of the list of all the observed sources (409). The full table is available at CDS via anonymous ftp.} 
\end{center}
\end{table*}

\section{Line selection }\label{ch:lines}

\subsection{\co}
As a rare isotopologue of carbon monoxide, \co\ is an excellent tracer of column density in star-forming regions. Chemically, it is relatively stable compared to other species. C$^{18}$O, like all carbon monoxide isotopologues, can be affected by depletion on dust grains surfaces, but this only occurs in the innermost, densest and lowest temperature cloud core regions (e.g., \citealt{caselli1999,bacmann2002,giannetti2014}). \co\ is, therefore, used as a reference molecule to measure the relative abundances of the other molecules in this paper.

\subsection{\hco, HOC$^{+}$ and HCO}
\hco\ is a rare isotopologue of HCO$^{+}$ (formylium) and mostly optically thin in molecular clouds. HCO$^{+}$ is a high-density tracer (critical density $\sim 10^{5}$\,cm$^{-3}$ for HCO$^{+}\,(J=1-0$) with the collisional rate at 20\,K taken from the Leiden Atomic and Molecular Database \citep{schoier2005}) that shows enhanced abundances in regions of higher fractional ionization and toward outflows where shock-generated radiation fields are present \citep{rawlings2000}. Hydroxy methylidyne (HOC$^{+}$) is a reactive ion that is almost exclusively related to regions with a high ionizing flux (either PDRs or X-ray-dominated regions; \citealt{fuente2003, rizzo2003,rizzo2005}). However, there are few reported detections of this rare molecule. HOC$^{+}$ has been detected toward the \uchii\ of Mon R2 together with the other reactive ion CO$^{+}$ \citep{rizzo2003}. The Formyl radical (HCO) has been mostly studied in PDRs toward the Orion Bar \citep{schilke2001}, the Horsehead nebula \citep{gerin2009}, and even the starburst galaxy M82 \citep{garcia-burillo2002}. In the Horsehead nebula, its emission strongly correlates with polycyclic aromatic hydrocarbon (PAH) and \cch\ emission, and its measured abundance reaches $X$(HCO) $\simeq 1-2\times10^{-9}$ \citep{gerin2009}. In addition, the HCO/\hco\ column density ratio in PDRs with \hii\ regions (e.g., \citealt{schenewerk1988,schilke2001,gerin2009}) shows higher values than in regions without \hii\ regions or any other signpost of star formation.  

\subsection{\cch\ and \hccch}
The ethynyl radical (\cch) and cyclopropynylidyne (\hccch) are small hydrocarbon species and well known for their association with PAH molecules \citep{rizzo2005}. Enhanced abundances of these small hydrocarbons were found in PDRs with intense UV fields \citep{fuente2003,teyssier2004,fuente2005,pety2005,rizzo2005,ginard2012}. Their spatial distribution follows PAH emission in general, but it is slightly different in extent \citep{pilleri2013,cuadrado2015}. In the PDRs of Mon R2 and the Orion Bar, the abundance of \cch\ is constant for a broad range of incident UV radiation strength, but the abundance of \hccch\ was appeared in usage of high in low UV PDRs \citep{cuadrado2015}. In addition, \cch\ shows higher column densities in high-UV irradiated PDRs, whereas column densities of \hccch\ are predicted to decrease for such PDRs \citep{cuadrado2015}.

\subsection{\hnc, \hcni, \hcnii, and CN}
Hydrogen cyanide (HCN) and hydrogen isocyanide (HNC) are used as tracers of dense gas within molecular clouds (e.g., \citealt{vasyunina2011}). The column density ratio HCN/HNC significantly depends on the temperature of the cloud as has been found in the Orion molecular cloud and several high-mass star-forming regions \citep{goldsmith1986,schilke1992,jin2015}. The cyanide radical (CN) at millimeter wavelengths is often used as a probe of dense gas and PDRs in the Galactic interstellar medium \citep{rodriguez-franco1998,boger2005}. Previous observational and theoretical studies have shown that CN abundances are enhanced in PDRs \citep{fuente1993,rodriguez-franco1998}. In particular, the observations toward the Orion Bar PDRs showed that the CN emission is located between the molecular ridge and the ionization fronts \citep{jansen1995,simon1997,rodriguez-franco1998}.

\section{Observations, source type, and data reduction}\label{ch:obs_data}
\subsection{Observations and source type}
The molecular line data were taken from unbiased spectral line surveys covering a frequency range of $\sim$84$-$115\,GHz (see \citealt{csengeri2016b} for details). They contain a number of molecular lines with different rotational energy levels. The line survey is observed with the Eight MIxer Receiver (EMIR)\footnote{http://www.iram.es/IRAMES/mainWiki/EmirforAstronomers} of the IRAM 30\,m telescope (Project IR: 181-10 and 037-12) toward 409 clumps selected from the ATLASGAL compact source catalog. The velocity resolution of these spectral line data is $\sim$ 0.5\,\kms. The largest beam size in the 3\,mm atmospheric window is $\sim$29$''$. We applied the forward efficiency ($\eta_{l}$ = 0.95) and the main-beam efficiency ($\eta_{\rm MB}$ = 0.81) to convert antenna temperatures, $T_{a}^{*}$, to the main beam brightness temperatures, $T_{\rm MB}$.
Here we divide the ATLASGAL sample into two groups, which are \hii\ and non-\hii\ regions, according to the presence of \hii\ regions based on detection of mm-RRL and radio continuum emission (see \citealt{kim2017} for details). Table \ref{tb:sou} displays information about the observed sources and the clump classification from \cite{urquhart2018}. This classification is used to further divide the group of non-\hii\ region sources into IR bright non-\hii\ and IR dark non-\hii\ regions.

\subsection{Data reduction}
\begin{figure}
\centering
\includegraphics[width=0.48\textwidth]{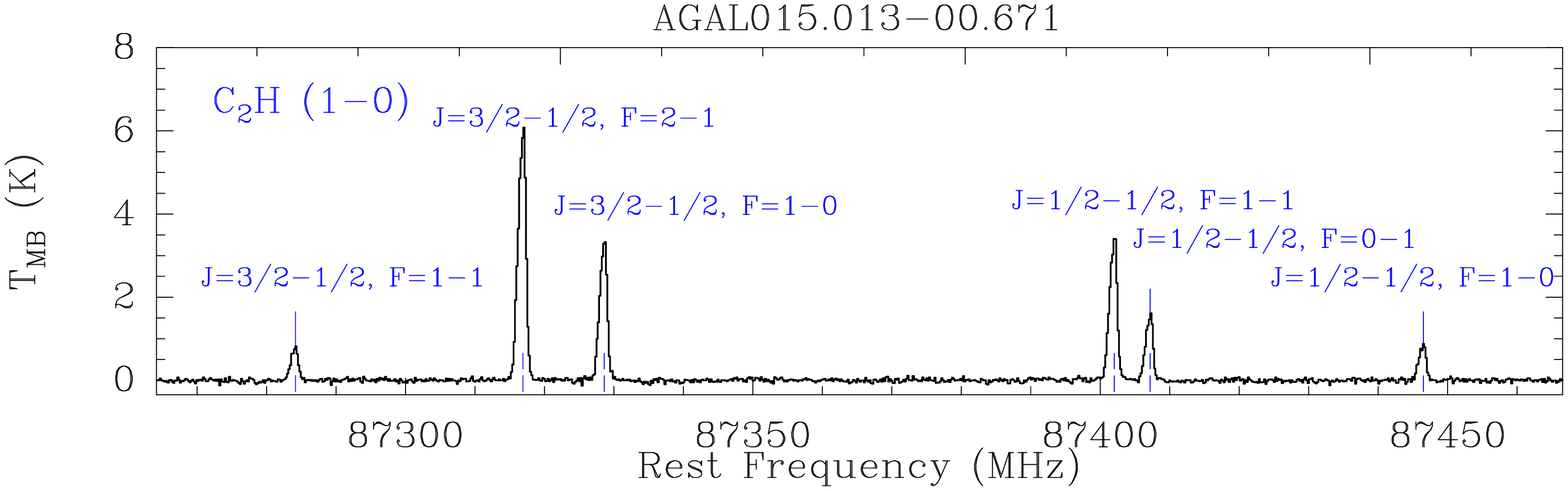}
\vskip 0.2cm
\includegraphics[width=0.48\textwidth]{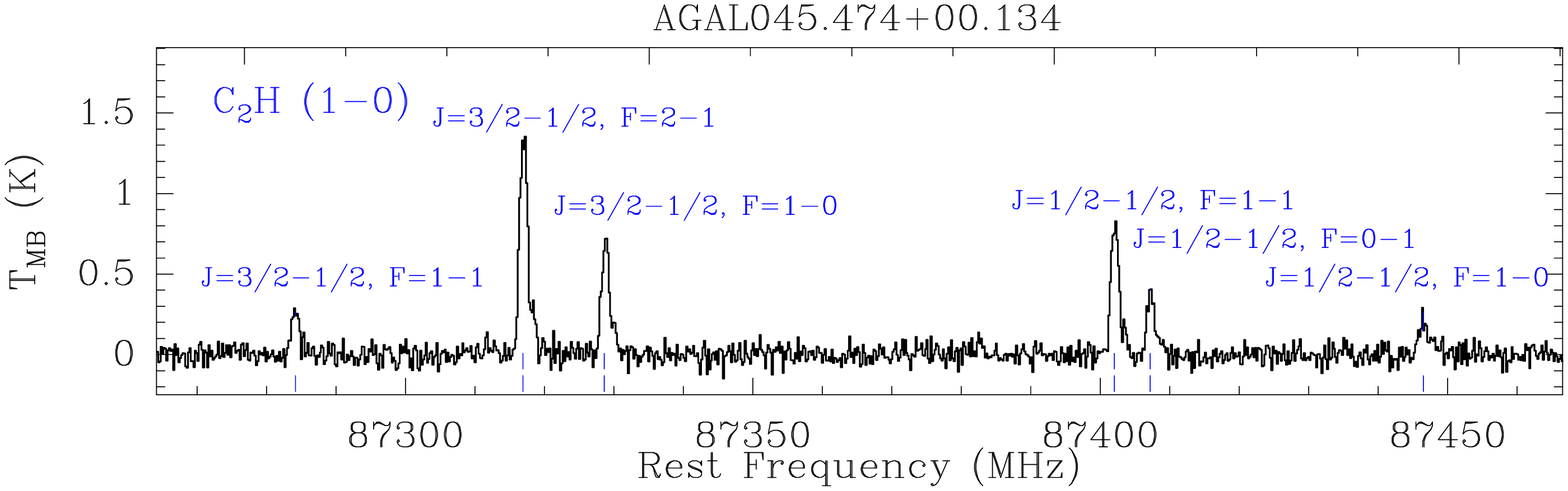}
\caption[]{\label{fig:spec_data1} Spectra of \cch\ hyperfine lines toward two example sources (AGAL015.013$-$00.671 and AGAL045.474$+$00.134). }
\end{figure}

\begin{figure}
\centering
\includegraphics[width=0.23\textwidth]{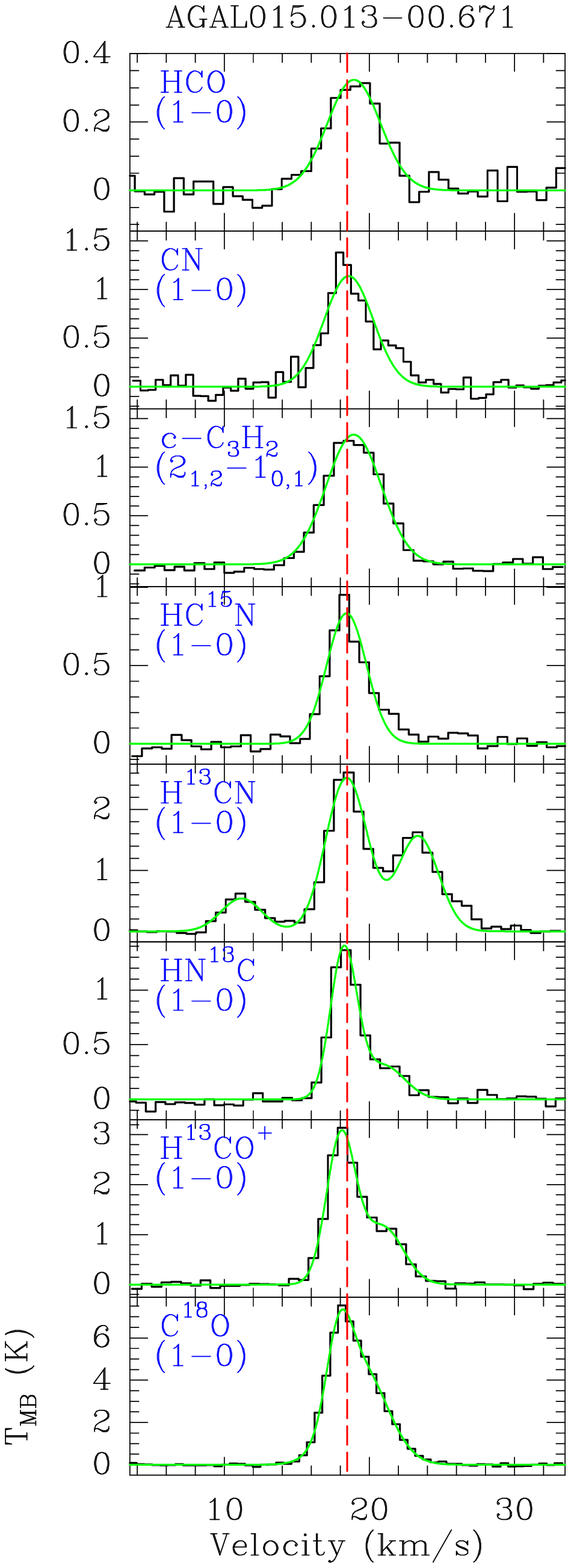}
\hspace{0.12cm}
\includegraphics[width=0.23\textwidth]{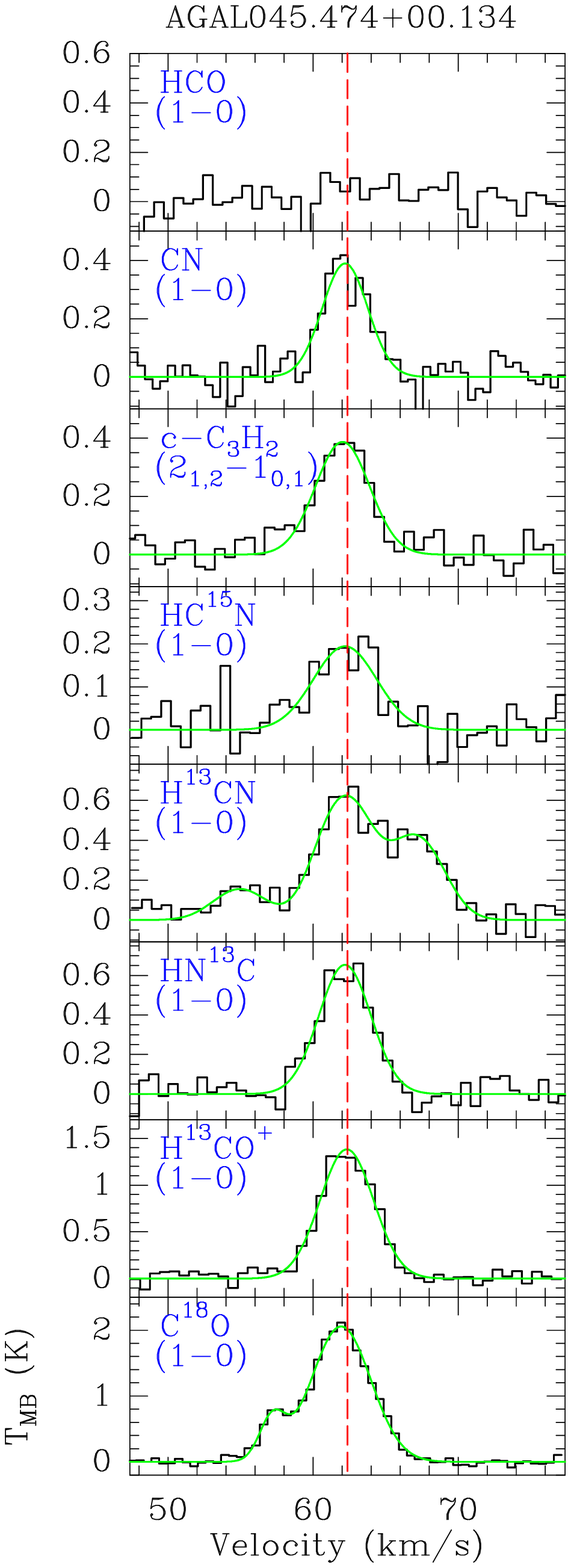}
\caption[]{\label{fig:spec_set2} From top to bottom, spectra of HCO, CN, \hccch, \hcnii, \hcni, \hnc, \hco, and \co\ toward the same example sources as in Fig.\,\ref{fig:spec_data1}. Hyperfine structure fitting was used for \hcni, and Gaussian profiles for the other molecular lines. The green lines represent Gaussian and HFS fitting results, and the red lines indicate the systemic velocity. HCO (1$-$0) was not detected toward AGAL045.474$+$00.134.} 
\end{figure}

We utilized the CLASS software\footnote{https://www.iram.fr/IRAMFR/GILDAS/doc/html/class-html/class.html} of the GILDAS package \citep{pety2005_gildas} for data reduction of the molecular line data and the WEEDS\footnote{https://www.iram.fr/IRAMFR/GILDAS/doc/html/weeds-html/node10.html} package within CLASS for line identification. The hyperfine structures of the C$_{2}$H ($N=1-0$) and H$^{13}$CN ($J=1-0$) transitions were fitted (see Figs. \ref{fig:spec_data1} and \ref{fig:spec_set2}) with the hyperfine structure (HFS) fit method\footnote{http://www.iram.es/IRAMES/otherDocuments/postscripts/classHFS.ps} of the CLASS software. The HFS fit method assumes that all hyperfine components share a single excitation temperature and local thermodynamic equilibrium (LTE). The HFS fit yields the radial velocity ($\varv_{\rm LSR}$), the line width ($\Delta \varv$) at the full-width of half maximum, and the total optical thickness ($\tau_{\rm tot}$). The line parameters of the other molecular lines were fitted with multi-Gaussian components if one Gaussian fit was insufficient.
\begin{table*}
\small
\centering
\caption{\label{tb:hfs_para}  The line parameters of hyperfine lines of \cch\ ($N = 1-0$) and \hcni\ ($J = 1-0$).}
\begin{tabular}{l l c c c c c}
\hline\hline
ATLASGAL name & Line &$\varv_{\rm LSR}$ & $\Delta \varv$ & $\tau_{\rm tot}$ & $T_{\rm MB}$ & rms \\
		 &       & (\kms) & (\kms) & & (K) & (K) \\
\hline\hline
AGAL008.671$-$00.356& \cch\ &   35.03$\pm$0.02~&    4.39$\pm$0.05~&    2.09$\pm$0.12~&3.14&    0.06\\
                    \hline
AGAL010.151$-$00.344& \cch\ &    9.55$\pm$0.01~&    5.10$\pm$0.02~&    0.54$\pm$0.03~&2.44&    0.04\\
					& \hcni\ &  9.70$\pm$0.16~&    4.93$\pm$0.37~&    3.05$\pm$0.64~&0.43&    0.04\\
                    \hline
AGAL010.168$-$00.362& \cch\ &   14.30$\pm$0.03~&    6.50$\pm$0.07~&    1.04$\pm$0.13~&1.78&    0.04\\
					& \hcni\ & 14.02$\pm$0.09~&    5.86$\pm$0.31~&    0.84$\pm$0.35~&0.68&    0.04\\
                    \hline
\end{tabular}
\tablefoot{ We only provide here the first three of the whole table which is available at the CDS via anonymous ftp. $T_{\rm MB}$ is the peak intensity of the brightest component of a given molecular hyperfine lines. The listed \cch\ peak intensity indicates  the transition $NJF = 1(3/2)2-0(1/2)1$, and the peak intensity of \hcni\ is the transition $JF = 12-01$.}
\end{table*}

\begin{table*}
\small
\centering
\caption{\label{tb:gauss_para} Gaussian line parameters of \co, HCO, \hccch, CN, \hcnii, and \hnc.}
\begin{tabular}{l l c c c c c}
\hline\hline 
ATLASGAL & Line & Area & $\varv_{\rm LSR}$ & $\Delta \varv$ & $T_{\rm MB}$ & rms  \\
 Name         &      & (K \kms) & (\kms)   & (\kms) & (K)  & (K)\\
\hline
AGAL008.671$-$00.356&HCO&	       0.83$\pm$0.10&   36.43$\pm$0.48&    6.57$\pm$0.78&    0.12&     0.03\\
					& \hco\ & 11.93$\pm$0.91&   34.20$\pm$0.06&    3.07$\pm$0.10&    3.65&    0.05\\
                    & \hccch\ &  5.40$\pm$0.33&   36.29$\pm$0.18&    5.73$\pm$0.26&    0.88&    0.05\\
                    & \hnc\ &  10.12$\pm$0.14&   35.25$\pm$0.03&    4.47$\pm$0.07&    2.13&    0.05\\
                    \hline
AGAL010.151$-$00.344& \co\ &17.65$\pm$0.09&    9.19$\pm$0.01&    5.48$\pm$0.03&    3.03&    0.03\\
					&HCO&      0.86$\pm$0.08&    8.70$\pm$0.36&    7.42$\pm$0.75&    0.10&    0.02\\	
                    & \hco\ &  7.98$\pm$0.11&    9.50$\pm$0.04&    5.26$\pm$0.09&    1.43&    0.04\\
                    & \hccch\ &  8.24$\pm$0.10&    9.48$\pm$0.03&    4.99$\pm$0.07&    1.55&    0.03\\
                    &CN&      5.06$\pm$0.16&    9.34$\pm$0.08&    5.18$\pm$0.19&    0.92&    0.06\\
                    & \hcnii\ &   0.35$\pm$0.09&   10.65$\pm$0.33&    2.99$\pm$1.07&    0.11&    0.04\\
                    & \hnc\ &   3.13$\pm$0.10&    9.49$\pm$0.07&    4.43$\pm$0.16&    0.66&    0.04\\
                    \hline
\end{tabular}
\tablefoot{For multiple velocity components, we only tabulate line parameters of the component detected in all molecular species. The full table is available at the CDS via anonymous ftp.}
\end{table*}


\section{Results}\label{ch:results}
\subsection{Line parameters}
Since the local standard of rest (LSR) velocities of the sources were unknown when the observations were carried out, they were observed with a velocity of 0\,\kms. Consequently, due to cases with a significantly different LSR velocity, the numbers of observed and detected lines in the sources vary slightly, and the molecules non-detected toward each source are given in Table \ref{tb:sou}. Tables \ref{tb:hfs_para} and \ref{tb:gauss_para} provide an example of the full tables of the HFS fitted, and Gaussian fitted line parameters, which are only available in electronic form. In cases where several velocity components were found ($\sim$ 1\%$-$8\% of the detected sources), we only selected a component with a common velocity in all detected lines. Figures \ref{fig:spec_data1} and \ref{fig:spec_set2} show the spectral lines and fitted line profiles of AGAL015.013$-$00.671 and AGAL045.474$+$00.134 as examples (see captions for details on transitions presented). 

\begin{figure}[ht]
\centering
\includegraphics[width=0.24\textwidth]{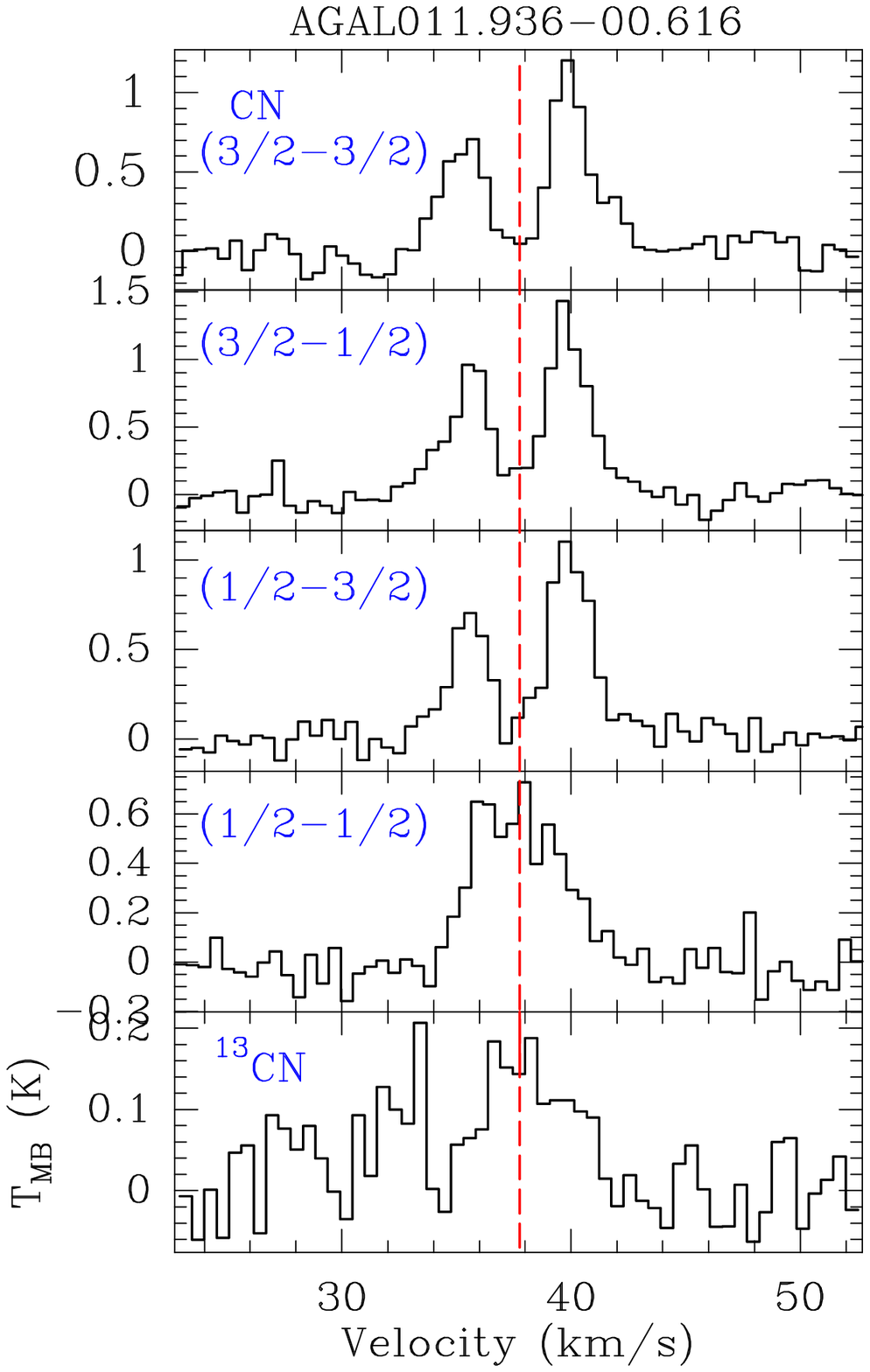}
\includegraphics[width=0.24\textwidth]{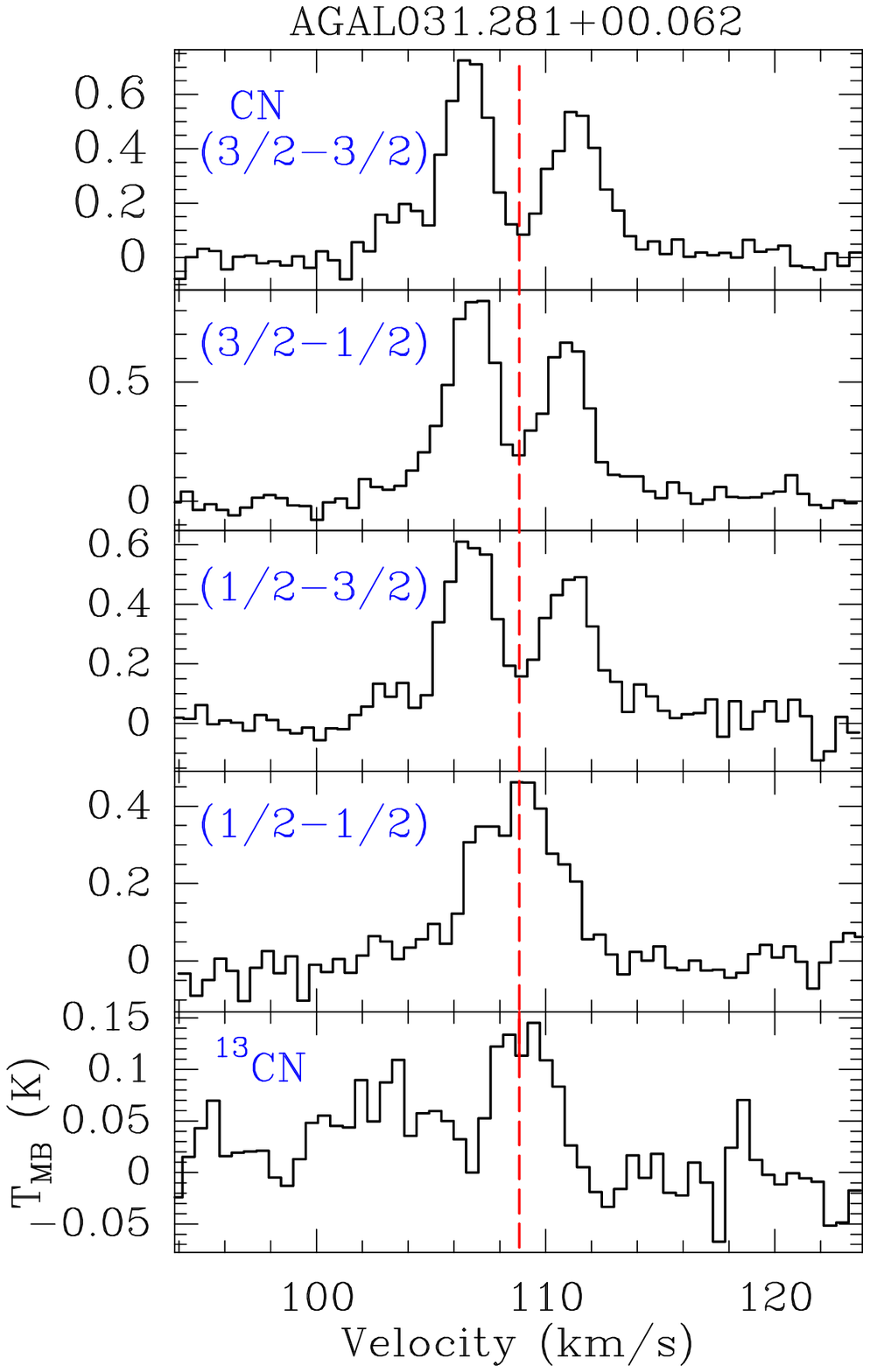}
\caption[]{\label{fig:cn_source} The self-absorption of CN lines of AGAL011.936$-$00.616 (left) and AGAL031.281$+$0.062 (right). Spectral lines of CN ($NJ= 1(1/2) - 0(1/2)$) and $^{13}$CN ($N= 1- 0, J=3/2-1/2, F_{1}= 2- 1, F= 3- 2$) transitions. The red vertical lines indicate the systemic velocities measured by \hco\ lines.}
\end{figure}
For CN and HCO, we performed a single Gaussian fit to the molecular lines although several of their hyperfine lines have been detected. CN hyperfine lines at 3\,mm wavelengths have substantial deviations from their relative intensities expected in LTE, making the hyperfine fits uncertain. In particular, this effect is stronger for the brightest three components than for the weakest component, which is considered to be optically thin. To reduce optical depth effects and to avoid non-LTE excitation, we used the weakest component of CN ($NJF = 1(1/2)3/2 - 0(1/2)3/2$) for all analysis presented in this work. In addition to these hyperfine anomalies, some sources show self-absorption in the three brightest components, which will be discussed in Section \ref{ch:cn_self-absorp}. The four HCO hyperfine lines are considerably weaker than the other molecular lines, and thus only a few sources have all the four HCO hyperfine components clearly detected. The brightest component ($NJF = 1(3/2)2 - 0(1/2)1$) is strong enough to secure a detection of the molecule. But to increase the signal-to-noise ratio (S/N), the detection rates and Gaussian fits of HCO are based on stacked spectra. The stacked HCO lines are scaled to the brightest component transition. For the stacking, each HFS component is weighted and scaled by its relative intensity. The stacking increased the S/N by a factor of 2. As shown in Fig. \ref{fig:spec_set2}, the three hyperfine components of \hcni\ observed toward the clumps hosting \hii\ regions are often blended. As a result, the measured line parameter of \hcni\ toward some sources are associated with high uncertainties, and we will discuss this in later sections. 

\begin{table*}
\centering
\small
\caption[]{\label{tb:detection} Detection rates of the observed molecules. }
\begin{tabular}{cccccccccccc}
\hline \hline
Molecule & \multicolumn{3}{c}{\co} && \multicolumn{3}{c}{HCO} && \multicolumn{3}{c}{\hco} \\ \cline {2 - 4} \cline{6-8} \cline{10-12}
Source type & All & \hii\ & non-\hii&& All & \hii\ & non-\hii&& All & \hii\ & non-\hii \\  \cline {2 - 4} \cline{6-8} \cline{10-12}
rms (K)   & 0.050 &  $-$ &  $-$ && 0.029 & $-$  &  $-$   && 0.041 & $-$  &  $-$    \\ 
Number of Observed source & 385 & 102 & 283 && 405 & 103 & 302 && 403 & 103 & 300 \\ 
Number of Detected source & 385 & 102 & 283 && 131 &  69 &  62 && 400 & 100 & 300 \\ 
Detection Rate (\%)       & 100 & 100 & 100 &&  32 &  66 &  20 &&  99 &  97 & 100 \\ 
\hline
Molecule & \multicolumn{3}{c}{\cch} && \multicolumn{3}{c}{ \hccch} && \multicolumn{3}{c}{CN} \\ \cline {2 - 4} \cline{6-8} \cline{10-12} 
Source type & All & \hii\ & non-\hii&& All & \hii\ & non-\hii&& All & \hii\ & non-\hii \\  \cline {2 - 4} \cline{6-8} \cline{10-12}
rms (K) & 0.039 & $-$ & $-$&& 0.038 & $-$ &$-$ && 0.069 &$-$  &$-$  \\ 
Number of Observed source & 403 & 102 & 301 && 404 & 102 & 302 && 386 & 102 & 284 \\ 
Number of Detected source & 399 & 102 & 297 && 403 & 102 & 301 && 363 & 102 & 261 \\ 
Detection Rate (\%)       &  99 & 100 &  98 &&  99 & 100 &  99 &&  94 & 100 &  91 \\ 
\hline
Molecule & \multicolumn{3}{c}{\hnc} && \multicolumn{3}{c}{\hcni} && \multicolumn{3}{c}{\hcnii} \\ \cline {2 - 4} \cline{6-8} \cline{10-12}
Source type & All & \hii\ & non-\hii&& All & \hii\ & non-\hii&& All & \hii\ & non-\hii \\  \cline {2 - 4} \cline{6-8} \cline{10-12}
rms (K) & 0.040 & $-$ &$-$ && 0.040 &$-$ &$-$ && 0.038 &$-$ &$-$  \\ 
Number of Observed source  & 405 & 103 & 302 && 403 & 101 & 302 && 405 & 103 & 302 \\ 
Numnber of Detected source & 384 &  94 & 290 && 305 &  94 & 211 && 161 &  82 & 79 \\ 
Detection Rate (\%)        &  94 &  91 &  96 &&  75 &  93 &  69 &&  39 &  79 & 26 \\ 
\hline
\end{tabular}
\tablefoot{For all observed sources, the rms of the line survey varies with observing frequency and is given for each molecule.} 
\end{table*}

\subsection{Self-absorption profiles of CN emission lines}\label{ch:cn_self-absorp}
Self-absorption features are found in the three strongest CN lines ($F=3/2-3/2,\,F=3/2-1/2$ and $F=1/2-1/2$) toward 25 \hii\ and 15 non-\hii\ region clumps. The features indicate the clumps have inhomogeneous structures in temperature and density, and, because of such structures, diffuse and low-temperature gas in the foreground absorbs the emitted radiation from CN molecules along the line of sight. Figure \ref{fig:cn_source} displays two example sources showing self-absorption features. These line profiles represent different kinds of gas motions within the clumps. All CN self-absorption spectral lines plots are available at CDS via anonymous ftp. In the left panel of Fig. \ref{fig:cn_source}, AGAL011.936$-$00.616 shows red-skewed emission features with self-absorption, which might be caused by the expansion of molecular cores \citep{smith2012} or outflows. On the contrary, we find blue-skewed emission features with self-absorption toward AGAL031.261$+$00.062 (right panel). Blue-skewed profiles indicate potential infall motions within clumps \citep{lucas1976,myers1996,wyrowski2016}. 

\begin{figure}
\centering
\includegraphics[width=0.48\textwidth]{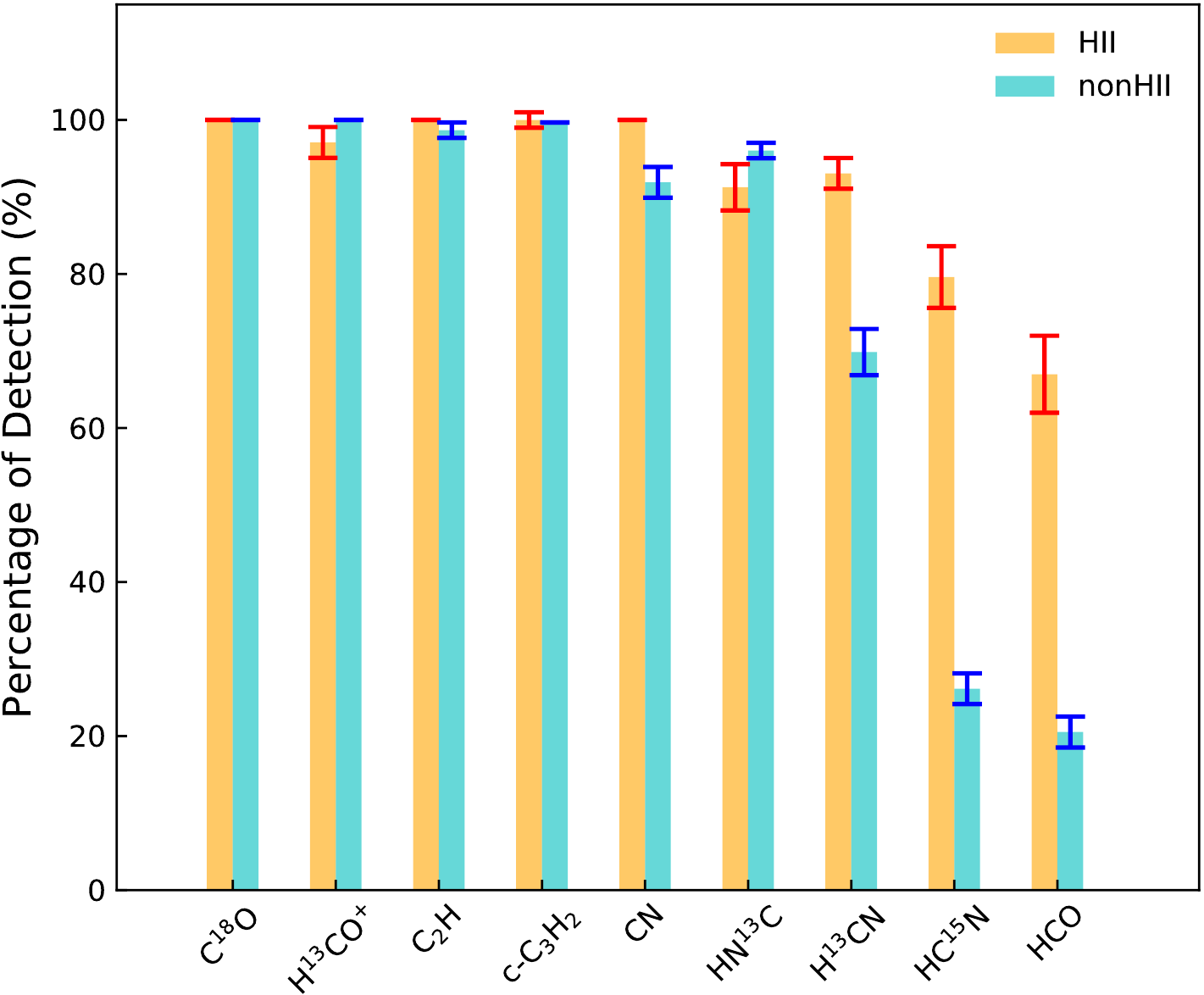}
\caption[]{\label{fig:detection} Detection rates of typical PDR tracer molecules and C$^{18}$O as a reference molecule. Binomial statistics was used for estimating uncertainties marked with the red and blue error bars.}
\end{figure}
\begin{figure}
\centering
\includegraphics[width=0.49\textwidth]{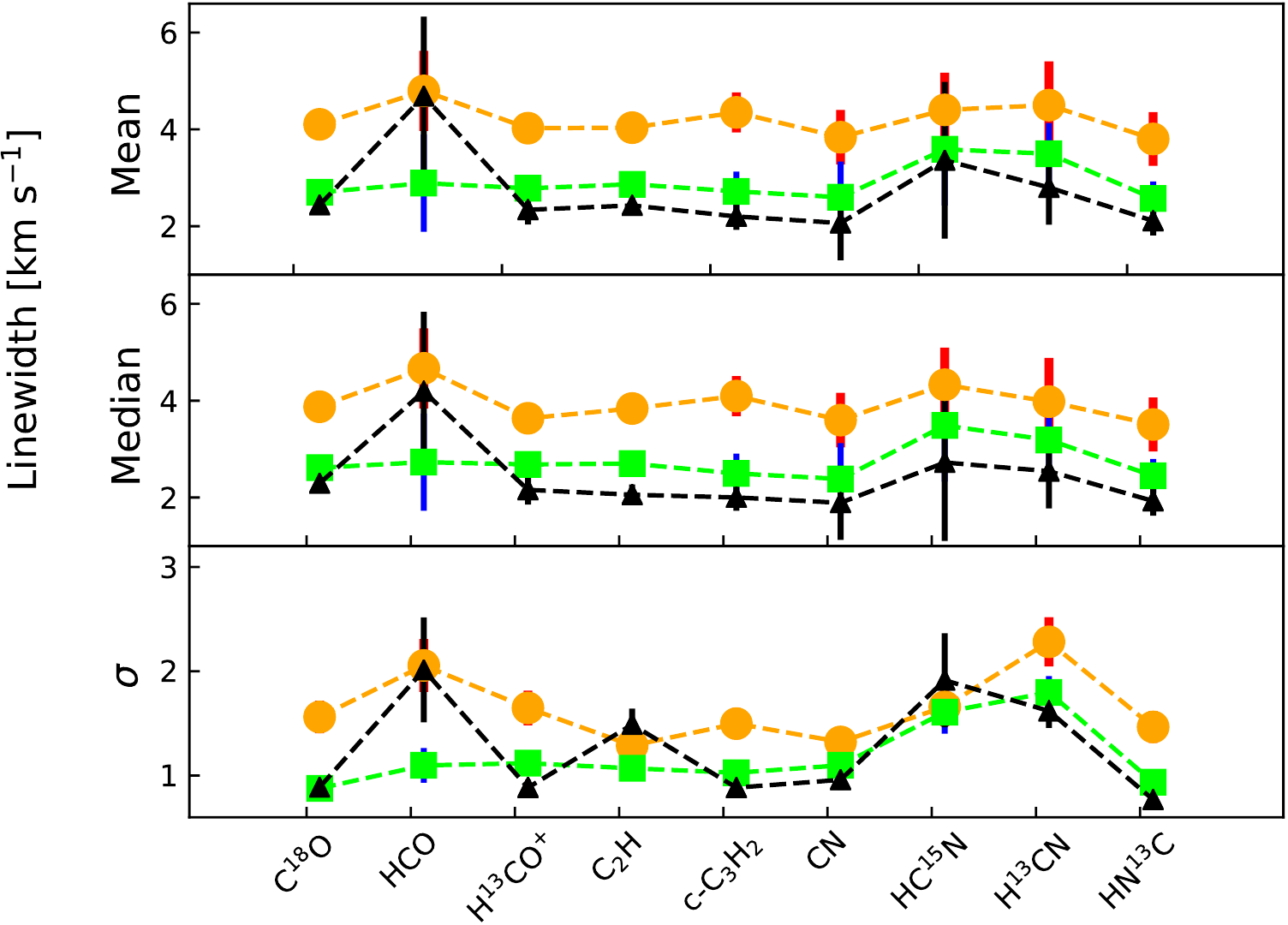}
\caption[]{\label{fig:linewdith_all} Mean, median, a standard deviation ($\sigma$) of line widths of the selected molecular lines. The $\sigma$ shows the source to source variation of the widths. The error bars for the mean and median values are based on typical Gaussian fit errors with 2$\sigma$. The error bars of the standard deviation of line widths is the standard deviation error. Different colors show three subgroups of our sample, which are \hii\ in orange, IR bright in green, and IR dark non-\hii\ regions in black}. 
\end{figure}
\begin{figure*}[ht]
\centering
\includegraphics[width=0.65\textwidth]{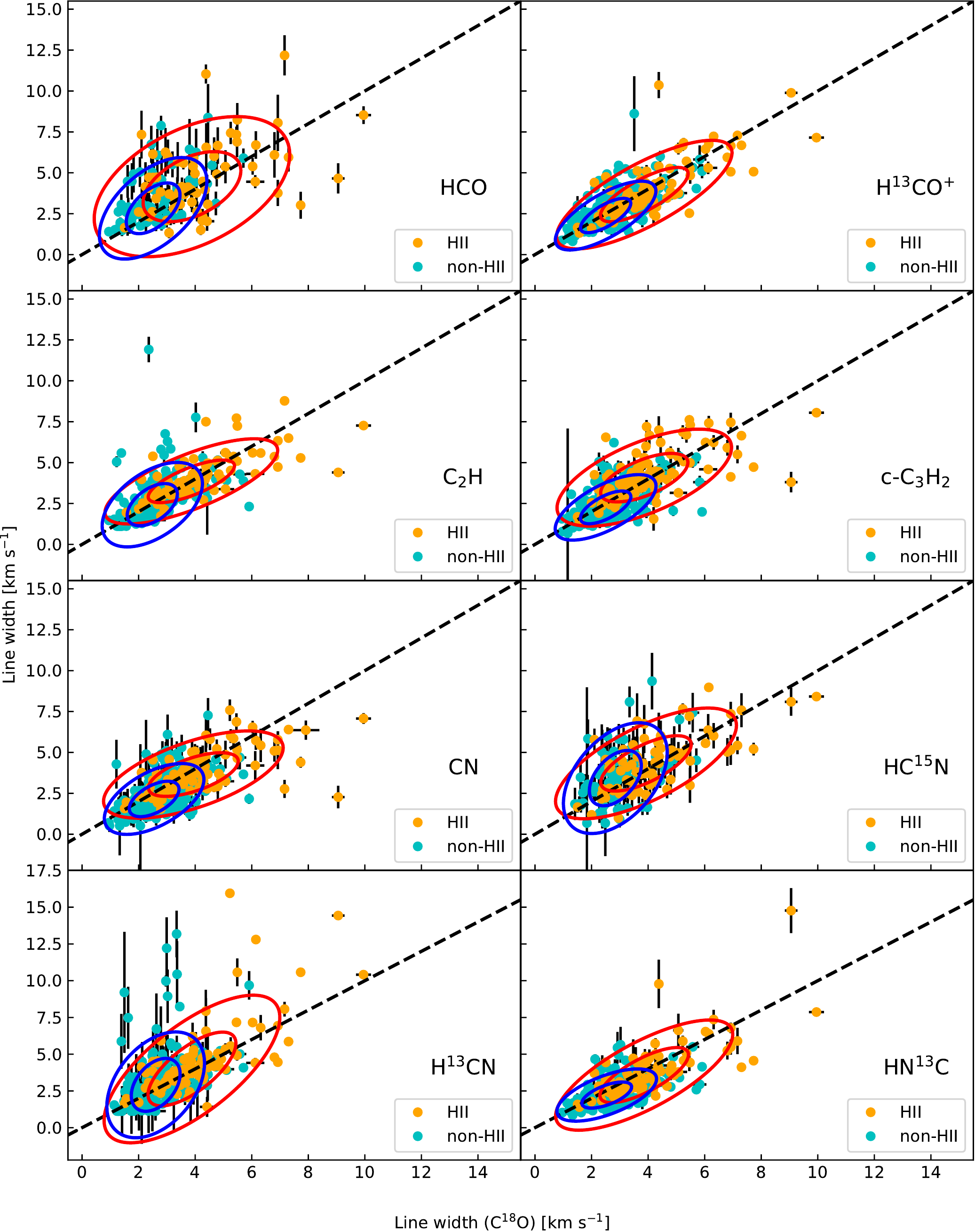}
\caption[]{\label{fig:linewidth} The orange circles indicate ATLASGAL clumps containing \hii\ regions and the cyan circles represent clumps without the presence of \hii\ regions, which consist of IR bright non-\hii\ and dark non-\hii\ regions. The black dashed lines indicate equality. The error bars are obtained from Gaussian or HFS fits. The red and blue ellipses are covariance error ellipses with 1$\sigma$ (inner ellipse) and 2$\sigma$ (outer ellipse) for the two groups.}
\end{figure*}
\subsection{Detection rates}

Table\,\ref{tb:detection} shows detection rates for the whole sample, including \hii\ and non-\hii\ regions. The detection of molecules was based on applying a 3\,$\sigma_{\rm rms}$ threshold. We detect \co\ lines toward all the observed clumps (100\%). This result shows that the ATLASGAL dust clumps trace the high column density parts of molecular clouds. In addition, we also find high detection rates for \hco\ (99\%), \cch\ (99\%), \hccch\ (99\%), CN (94\%), and \hnc\ (94\%). We find a reasonably high detection rate for \hcni\ (75\%). Even the rare isotopologue \hcnii\ is detected with a rate of 39\%. However, the other cation, HOC$^{+}$ was detected in only five clumps (below 3\,$\sigma$  with an rms $\sim$ 0.041\,K), which are marked in Table\,\ref{tb:sou}. It is probably due to its low abundance (including unknown beam dilution). HOC$^{+}$ fractional abundances toward typical PDRs, including the Orion Bar and Mon R2, are in a range of $\sim 9\times10^{-11} - 7\times10^{-13}$ \citep{savage2004}. In addition, its large dipole moment and $A_{\rm ij}$ imply the critical density of this molecule is probably higher than others with small dipole moment if collisional rates are similar. These tentative detections of HOC$^{+}$ are excluded from the following statistical analyses.

Figure \ref{fig:detection} shows the detection rates of the nine molecules toward \hii\ (orange bars) and non-\hii\ (cyan bars) region sources. Differences in detection rates for some molecular transitions (i.e., \hcni, \hcnii, and HCO) toward the two groups are distinct, with the \hii\ regions showing higher detection rates that can be related to higher column densities.
According to \KS\ tests and median values of \hh\ column densities ($N$(\hh)) for the ATLASGAL clumps, the median value of the $N$(\hh) determined for the \hii\ region sources is not significantly different from the value found for non-\hii\ region sources. The $N$(\hh) are taken from \cite{konig2017}, and \cite{urquhart2018} and were derived by analysing dust continuum emission with adopting dust opacity of 1.85 cm$^{2}$\,g$^{-1}$ and $\mu_{\rm H_{2}}$ of 2.8 that is mean molecular weight of the interstellar gas for a hydrogen molecule. The $N$(\hh) values of both \hii\ and non-\hii\ regions are mostly in a range of $\sim 4-6\times10^{22}$\,cm$^{-2}$ (for \hii\ regions, $1.14\times10^{22}$\,cm$^{-2}\leq$\,\,$N$(\hh)\,\,$\leq$\,\,$1.06\times10^{24}$\,cm$^{-2}$ compared to $1.25\times10^{22}$\,cm$^{-2}\leq$\,\,$N$(\hh)\,\,$\leq$\,$3.43\times10^{23}$\,cm$^{-2}$ for non-\hii\ regions). Another possibility is a distance effect leading to weaker molecular lines for more distant sources. However, this can be excluded because of the detection rates for all the molecules for a distance limited sample (3 $-$ 5\,kpc) are approximately the same (see Fig.\,\ref{appendix:dr_dist}).

\begin{figure*}[ht]
\centering
\includegraphics[width=0.33\textwidth]{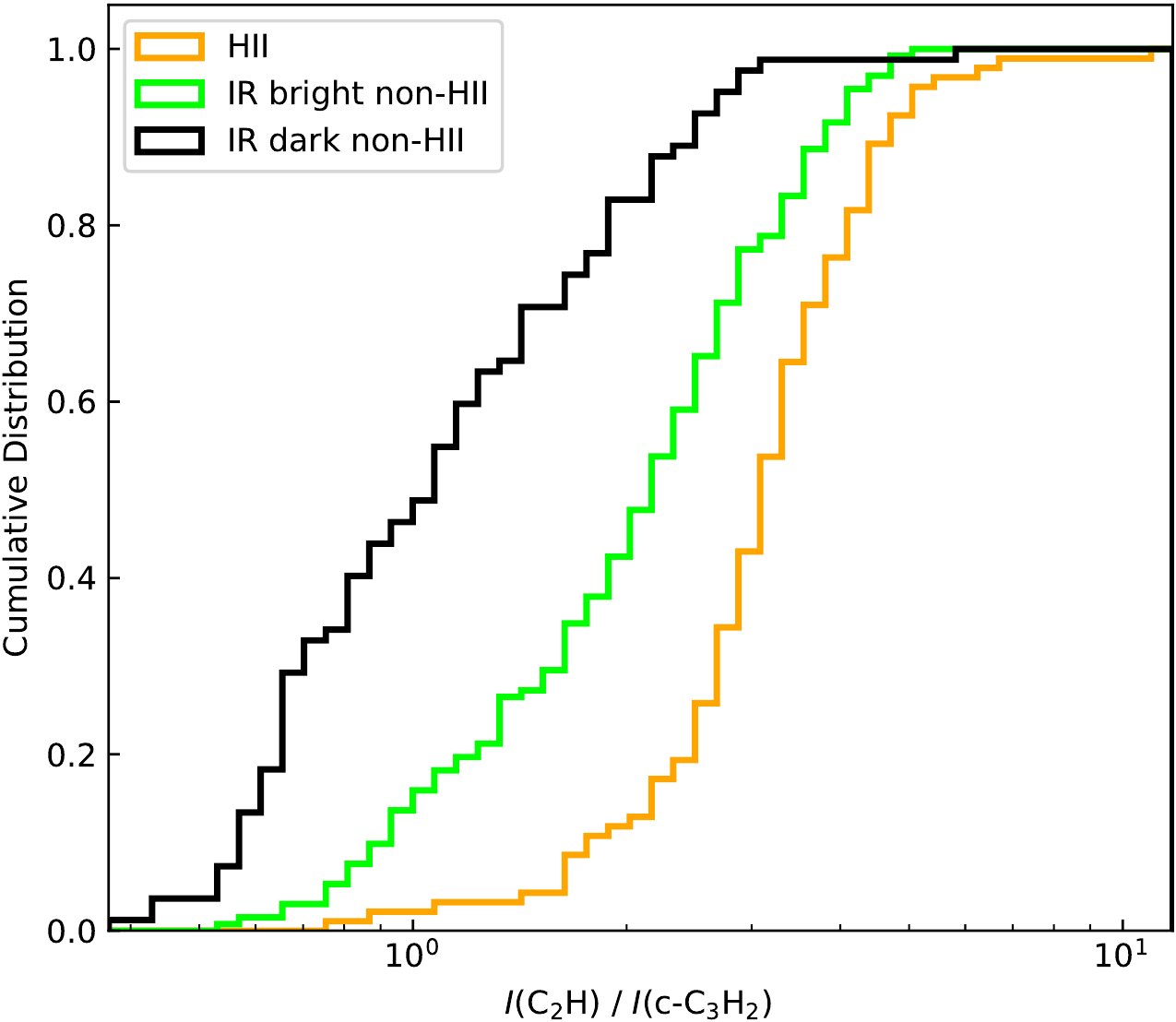}
\includegraphics[width=0.33\textwidth]{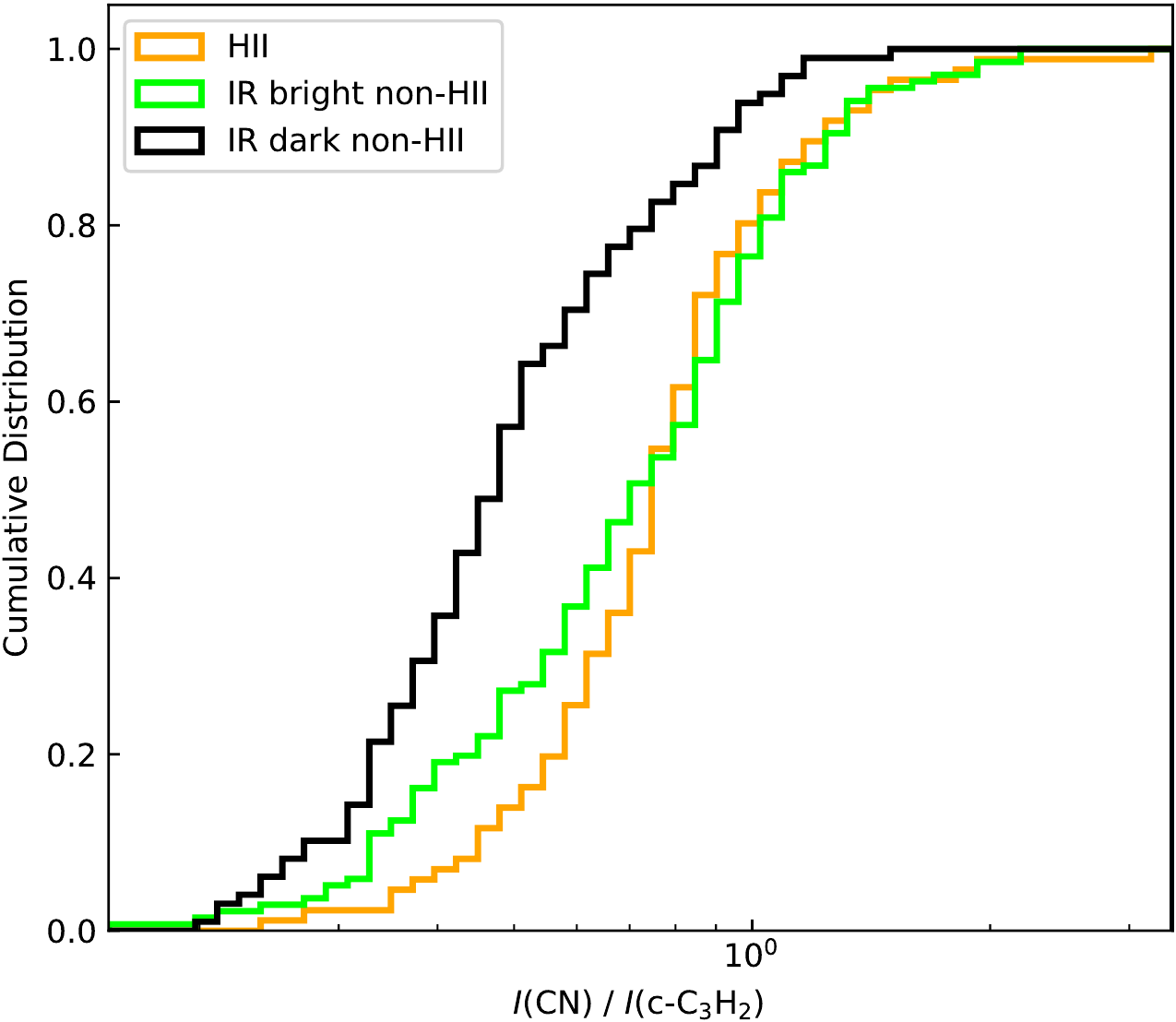}
\includegraphics[width=0.33\textwidth]{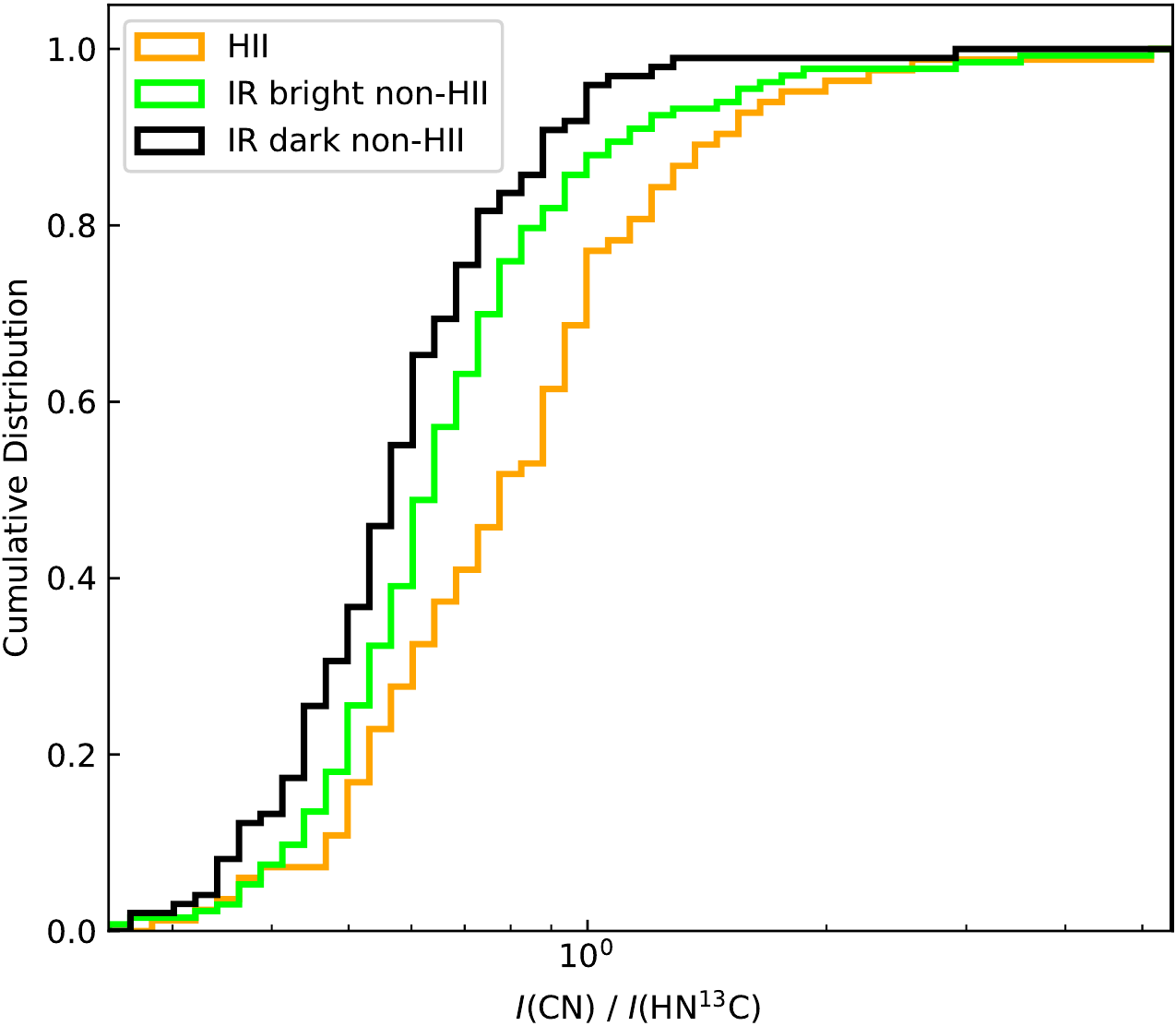}
\caption{\label{fig:int_ratio} Cumulative distribution plots of integrated intensity ratios of \cch\ to \hccch\ (left panel) and CN to \hccch\ (middle panel) and to \hnc\ (right panel). The orange curves are \hii\ regions. The green and black curves indicate IR bright non-\hii\ and IR dark non-\hii\ regions.  }
\end{figure*}
\begin{table*}
\small
\centering
\caption{\label{tb:col_den} Column density of molecules in units of cm$^{-2}$. }
\begin{tabular}{c c c c c c c c c c }
\hline \hline
ATLASGAL & \co\ & HCO & \hco\ & \cch\ & \hccch\ & CN & \hcnii\ & \hcni\ & \hnc\ \\
Name & $\times10^{16}$ & $\times10^{13}$ & $\times10^{13}$ & $\times10^{15}$ & $\times10^{13}$ & $\times10^{15}$ & $\times10^{13}$ & $\times10^{13}$ &$\times10^{14}$ \\
\hline
AGAL006.216$-$00.609& 1.08 & $\cdots$ & 0.41 & 0.79 & 2.56 & 0.86 & 0.05 & 0.67 & 0.07\\
AGAL008.049$-$00.242& 0.37 & $\cdots$ & 0.08 & 0.12 & 0.57 & 0.23 & $\cdots$ & 0.16 & 0.02 \\
AGAL008.671$-$00.356& $\cdots$ & 2.12 & 0.96 & 1.73 & 3.67 & $\cdots$ & $\cdots$ & $\cdots$ & 0.17  \\
AGAL008.684$-$00.367& 2.90 & $\cdots$ & 1.00 & $\cdots$ & 2.42 & 1.31 & 0.21 & 3.37 & 0.13 \\
AGAL008.706$-$00.414& 1.26 & $\cdots$ & 0.31 & 0.42 & 1.81 & 0.61 & $\cdots$ & 0.98 & 0.06 \\
\hline
\end{tabular}
\tablefoot{This table shows a portion of the full table that is available at the CDS via anonymous ftp. Not all lines were observed in all sources, and thus some column densities are missing in Table \ref{tb:col_den} (see Table \ref{tb:sou} for missing observational data). }
\end{table*}

\subsection{Line widths and integrated fluxes}
To investigate whether different molecular lines are originating from the same gas or show different amounts of turbulence, we compared the mean, median, and standard deviations of the measured line widths. Figure \ref{fig:linewdith_all} shows these values with their uncertainties. The molecular line widths increase from less evolved clumps to more evolved clumps. Such a trend, broader molecular lines detected toward clumps in evolved stages, was also found in ammonia \citep{wienen2012,urquhart2013b} and \hco\ \citep{kim2017} data. The HCO line widths have considerable higher fit uncertainties due to the low line intensities (or lower S/N). Nevertheless, comparing the IR bright non-\hii\ and \hii\ regions, which have small fitting uncertainty, we find that broad HCO lines are often associated with \hii\ regions. The HCO, \hcni, and \hcnii\ lines observed toward IR dark non-\hii\ regions seem to be broader than the other molecular lines, but their errors are also obviously larger than those of the other source groups. The third panel from the top of Fig.\,\ref{fig:linewdith_all} exhibits the standard deviations ($\sigma$) of line width and its associated uncertainty. Except for \cch, CN, and \hcnii, the other molecular lines have broader distributions toward the \hii\ region sources compared with those toward non-\hii\ region sources.

Figure \ref{fig:linewidth} shows scatter plots comparing line widths of the lines from selected molecules with \co\ line widths. Covariance error ellipses are overlapped over the data points to visualize two-dimensional Gaussian distributed data\footnote{In the two-dimensional case, $1\sigma$ and $2\sigma$ show 39.4\,\% and 86.5\,\% confidences, respectively.}. The ellipses and data points of \hco\ and \co\ display pronounced correlations and well align on the equality line. On the other hand, the other molecular lines show some deviations from the equality lines. Some sources have remarkably broader line widths in \hcni\ with small uncertainty. Such broadening can be a result of high optical depth, leading to a blending of the HFS lines. If any internal turbulence within clumps causes these broad line widths, we expect to find such a trend also in \hcnii. But no sources with notably broader line widths of \hcnii\ are found.

To identify similarities in the line widths between the clump groups, we performed KS tests for the molecular line widths. The null hypotheses are rejected with small $p-$values less than 3$\sigma$ ( p-value $\ll 0.0013$, i.e., the confidence of $>$ 99.87\%). This means that statistically the line widths are significantly broader for the \hii\ regions than for the non-\hii\ regions, consistent with the results in Fig.\,\ref{fig:linewdith_all}. One possible cause for this may be distance bias. To test this, we compared molecular line widths and distances of the clumps (see Fig.\,\ref{appendix:w_dist}), but we do not find significant correlations. The distance effect cannot, therefore, be the main reason for the broad line widths of the \hii\ region group. For the \hii\ region group, we compared the line widths of molecular lines with RRL line widths. If the molecular lines are mainly emitted in PDRs, we might expect to find some correlations between them due to the dynamical interaction between ionized and photodissociated gas. However, we could not find any significant correlations.

Integrated intensity ratios of molecular lines have been used as a chemical clock tracing different evolutionary sequences of high-mass stars in molecular clumps (e.g., \citealt{rathborne2016,urquhart2019}). With the assumption that observed lines are optically thin, integrated intensity ratios gives us approximate abundance ratios of molecular lines compared with each other. Figure\,\ref{fig:int_ratio} shows cumulative distributions of the integrated intensity ratios of \cch, \hccch, and CN (i.e., \cch/\hccch\ in the left panel, CN/\hccch\ in the middle panel, and CN/\hnc\ in the right panel). We performed KS tests for the ratio differences among the three groups and found that we can reject the null hypotheses for the similarities of their ratios with small $p-$values $\ll$ 3$\sigma$. The \hii\ regions show brighter ratios of those molecular lines than the other two groups. The well separated cumulative distributions of \cch/\hccch\ show that the relative intensities of the small hydrocarbons increase with the evolution of high-mass clumps. This trend is also found for ATLASGAL sources in the southern hemisphere \citep{urquhart2019}.

\section{Analysis}\label{ch:analysis}
\begin{figure*}[ht]
\centering
\includegraphics[width=0.7\textwidth]{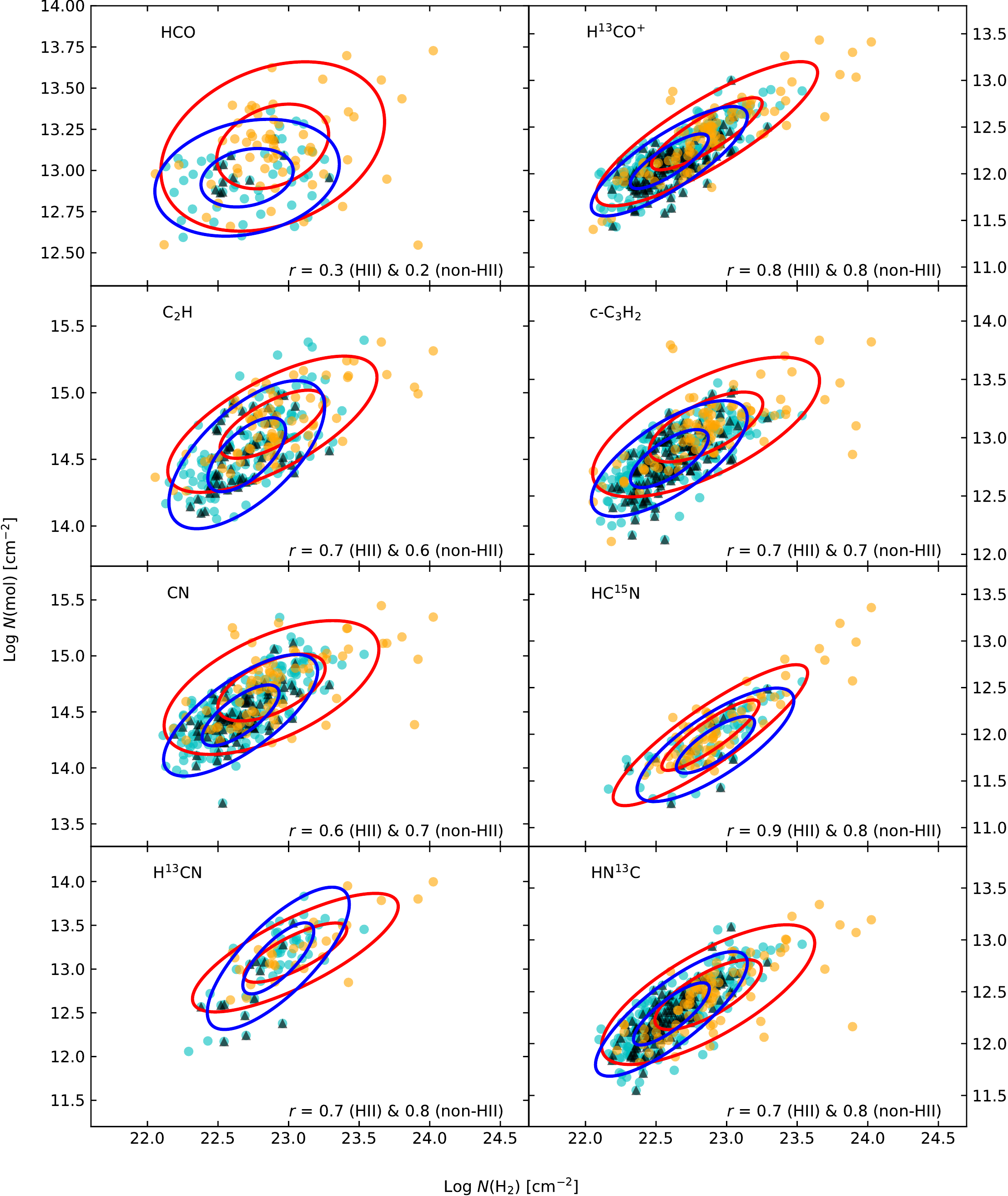}
\caption[]{\label{fig:n_n2} Column density of a given molecule as a function of \hh\ column density toward \hii\ and non-\hii\ regions with IR dark non-\hii\ regions superposed with black triangles. The ellipses show similar confidence contours as in Fig.\,\ref{fig:linewidth}. The $r$ parameter indicates Pearson correlation coefficients for \hii\ and non-\hii\ regions. }
\end{figure*}

\subsection{Column densities}
To compare abundances of the selected molecules with respect to \hh\ for different environments in dust clumps, as the first step, we estimated column densities of the molecules. Since the observed data only contain a single transition per molecule, several assumptions are required to derive column densities ($N$): 1) the molecular lines are emitted under LTE condition because the critical densities of the observed molecular transitions are lower than the \hh\ densities ($> 10^{5}$\,cm$^{-3}$) in high-mass star-forming regions; 2) at these high densities, the gas and dust temperatures are in equilibrium; 3) the observed line emission is considered to be optically thin.
For optically thin emission, the total column density is given by,
\begin{equation}
\begin{aligned}
\label{eq:optically_thin}
& N_{\rm tot}^{\rm thin} = \left(\frac{8\pi\nu^{3}}{c^{3}A_{ul}} \right)\left(\frac{Q(T)}{g_{u}} \right)\frac{{\rm exp} \left(\frac{E_{u}}{k_{\rm B}T_{\rm ex}} \right)}{{\rm exp}\left( \frac{h\nu}{k_{\rm B}T_{\rm ex}}\right) -1} \\ &
~~~~~~~~~~~~~~~~~~~~~~~~~~~~~~~~~~\times \frac{1}{[J_{\nu}(T_{\rm ex}) - J_{\nu}(T_{\rm bg})]} \int{\frac{T_{\rm MB}}{f}}\,d{\it \varv},
\end{aligned}
\end{equation}

\noindent  where $\nu$ is the frequency of a selected molecular transition, $A_{ul}$ is the Einstein coefficient for spontaneous emission, and $g_{u}$ is the statistical weight of the upper state level. $Q(T)$ and $k_{\rm B}$ are the partition function and Boltzmann constant, respectively. $E_{\rm u}$ is the energy of the upper level of the selected transition. $T_{\rm MB}$ is a main beam temperature of an observed source, and $f$ is the beam filling factor, which is the fraction of the beam filled by the source. Here we consider calculated column densities are beam-averaged values, and the medium is spatially homogeneous and larger than the size of the beam. These assumptions return the beam filling factor as 1. $J_{\nu}(T)$ is the Rayleigh-Jeans temperature, $J_{\nu}(T) \equiv \frac{h\nu/k_{\rm B}}{{\rm exp}(h\nu/k_{\rm B}T) - 1}$. $T_{\rm ex}$ is the excitation temperature which we approximate with the dust temperature, and $T_{\rm bg}$ is the background emission temperature assumed to be 2.7\,K that is the cosmic microwave background radiation. 

Since the optical depths of the \cch\ and \hcni\ lines were obtained by HFS fitting, correct column densities of these lines were estimated by multiplying Eq.\,(\ref{eq:optically_thin}) by a factor of $\tau/(1-e^{-\tau})$ for the non-optically thin case. For the other molecular lines, we utilized the optically thin case of Eq.\,(\ref{eq:optically_thin}), which gives us a lower limit of column density in case the lines become optically thick. We note that in some cases we failed to measure a $\tau_{\rm tot}$ with a fitting uncertainty below 50\%, either because some of \cch\ and \hcni\ hyperfine components are not separated sufficiently (e.g., the \hcni\ spectral line of AGAL045.474$+$00.134 in Fig.\,\ref{fig:spec_set2}) or their relative intensities have substantial deviations from the predicted LTE values. Consequently, we only estimated column densities of these molecules using their optical thickness when the uncertainty of the fit was smaller than 50\,\%. For the remaining sources with high $\tau_{\rm tot}$ uncertainties, we calculated their column densities as lower limits using the optically thin approximation (Eq.\,(\ref{eq:optically_thin})) and their values in Table \ref{tb:col_den}, but do not use them for following statistical analysis. The dust temperatures used in the analysis were determined by \cite{urquhart2018} using the method for the dust temperature determination from \cite{konig2017}. We estimated the partition function for an individual source and molecule by interpolating the data provided by the Cologne Database for Molecular Spectroscopy (CDMS, \citealt{muller2001}) and the Jet Propulsion Laboratory (JPL, \citealt{pickett1998}) line databases for a given temperature. The partition functions obtained from CDMS and JPL take into account the hyperfine splitting and ortho- and para-transitions. Overall, our column density measurements agree with those reported in previous studies (e.g., \citealt{sanhueza2012, gerner2014}). However, we find no clear differences between the sources representative of different evolutionary stages. We used simplifying assumptions such as the optically thin case, equating the dust temperature and gas temperature, and neglecting some factors like size of the emitting region. The measured column densities of a portion of all the observed clumps are tabulated in Table \ref{tb:col_den}.

\begin{figure}
\centering
\includegraphics[width=0.48\textwidth]{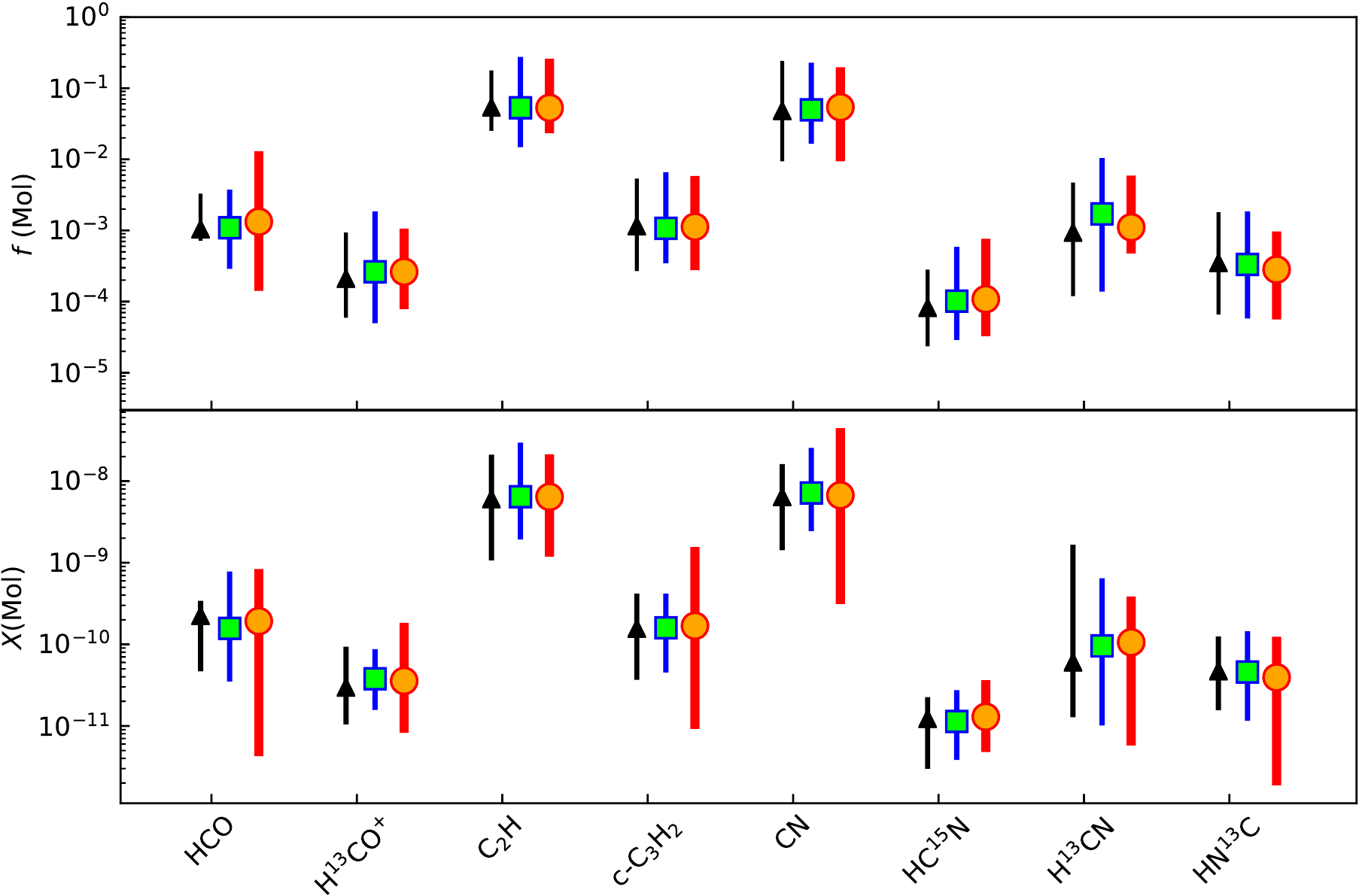}
\caption[]{\label{fig:x_mol} The abundance of the selected molecular species. $Upper$: Relative abundances ($f$) normalized by \co\ column density. $Lower$: Fractional abundances ($X$) with respect to \hh. The red circles, cyan squares, and black triangles indicate \hii, IR bright non-\hii, and IR dark non-\hii\ regions, respectively. The error bars show full ranges of fractional abundances for each molecule.}
\end{figure}
\begin{table*}
\centering
\caption{\label{tb:Nmol} Column densities ($N$) and fractional abundances ($X$ and $f$) relative to \hh\ and \co\ for \hii, IR bright and dark non-\hii\ regions in $a(x) = a\times10^{x}$. These values are median values for a given molecule. }
\begin{tabular}{l c c c c c c c c c c c}
\hline\hline
Molecule & \multicolumn{3}{c}{\hii}& &\multicolumn{3}{c}{IR bright non-\hii}& & \multicolumn{3}{c}{IR dark non-\hii} \\ \cline{2-4} \cline{6-8} \cline{10-12 }
 & $N$ [cm$^{-2}$] & $X$ & $f$ && $N$ [cm$^{-2}$] & $X$  & $f$ && $N$ [cm$^{-2}$] & $X$ & $f$ \\
\hline
\co\    & 9.5(15) & 1.2($-$07) & $-$ && 6.5(15) & 1.5($-$07) &$-$&& 5.9(15) & 1.4($-$07) & $-$\\
HCO     & 1.4(13) & 1.9($-$10) & 1.3($-$03) && 8.9(12) & 1.6($-$10) & 1.1($-$03) && 9.0(12) & 2.2($-$10) & 1.0($-$03)\\
\hco\   & 2.7(12) & 3.6($-$11) & 2.6($-$04) && 1.5(12) & 3.8($-$11) & 2.6($-$04) && 1.2(12) & 3.0($-$11) & 2.1($-$04) \\
\cch\   & 5.8(14) & 6.4($-$09) & 5.3($-$02) && 3.5(14) & 6.4($-$09) & 5.3($-$02) && 3.4(14) & 6.0($-$09) & 5.4($-$02) \\
\hccch\ & 1.2(13) & 1.7($-$10) & 1.1($-$03) && 7.0(12) & 1.6($-$10) & 1.1($-$03) && 6.4(12) & 1.6($-$10) & 1.1($-$03) \\
CN      & 5.2(14) & 6.7($-$09) & 5.4($-$02) && 3.2(14) & 7.2($-$09) & 5.0($-$02) && 2.7(14) & 6.4($-$09) & 4.8($-$02)\\
\hcnii\ & 9.8(11) & 1.3($-$11) & 1.1($-$04) && 8.7(11) & 1.1($-$11) & 1.0($-$04) && 5.6(11) & 1.2($-$11) & 8.2($-$05) \\ 
\hcni\  & 1.5(13) & 1.1($-$10) & 1.1($-$03) && 1.6(13) & 9.6($-$11) & 1.7($-$03) && 4.3(12) & 6.0($-$11) & 9.3($-$04) \\
\hnc\   & 3.0(12) & 3.9($-$11) & 2.8($-$04) && 2.0(12) & 4.6($-$11) & 3.3($-$04) && 1.8(12) & 4.7($-$11) & 3.5($-$04) \\
\hline
\end{tabular}
\end{table*}

Figure \ref{fig:n_n2} shows scatter plots of column density of \hh\ versus that of selected molecules. \hco\ and \hcnii\ exhibit an excellent correlation (Pearson correlation coefficient, $r$, $\geq 0.8$ with $p-$value $\ll$0.0013) for both the \hii\ and non-\hii\ region groups. In addition, \cch, \hccch, CN, \hcni\ and \hnc\ also show a good correlation ($r \sim$ 0.6 - 0.7 with $p-$values $\ll$0.0013). On the other hand, the $N$(HCO) has poor correlations for both groups ($r$ = 0.3 for \hii\ and $r$ = 0.2 for non-\hii\ regions) with $N$(\hh). HCO has been found to be enhanced in PDRs rather than in cold, dense molecular regions toward both Galactic objects \citep{schilke2001,gerin2009}, and extragalactic sources \citep{garcia-burillo2002}. The weak correlation supports the hypothesis that, where detected, the HCO emission is associated with PDRs on surfaces of the clumps and not their colder material. The column densities of \hco, \hccch, CN, and \hnc\ for the two groups are significantly different from each other (KS tests gives $p-$values $\ll$ 0.0013).

For HCO, \hcni\, and \hcnii, we cannot find significant differences between these groups according to KS tests that yield $p-$values $\gtrsim 0.004$. \cite{sanhueza2012} also did not find any increasing trend for $N$(\cch) with evolutionary stages. In our results, $N$(\hnc) is higher toward \hii\ regions, but this trend was not apparent in the results of \cite{sanhueza2012}. For comparison, \cite{miettinen2020} found some evidence that the C2H abundance decreases with clump evolution. Overall, the column densities of all molecules are higher in \hii\ regions than in the other groups.

\subsection{Molecular abundances}\label{sec:abundance}
Figure \ref{fig:x_mol} shows abundances relative to \co, $f$(mol), and fractional abundances relative to \hh, $X$(mol), of the selected molecules. The symbols and error bars represent median values, and full ranges of the molecular abundances, respectively. The median abundances are listed in Table \ref{tb:Nmol}. In Fig.\,\ref{fig:x_mol}, there are no significant abundance differences among the three source groups. When considering the uncertainties in the calculations, these results are not significantly different from the results of \cite{gerner2014}, who assumed higher temperatures for evolved sources such as \hii\ regions. 

\begin{figure*}[!ht]
\centering
\includegraphics[width=0.7\textwidth]{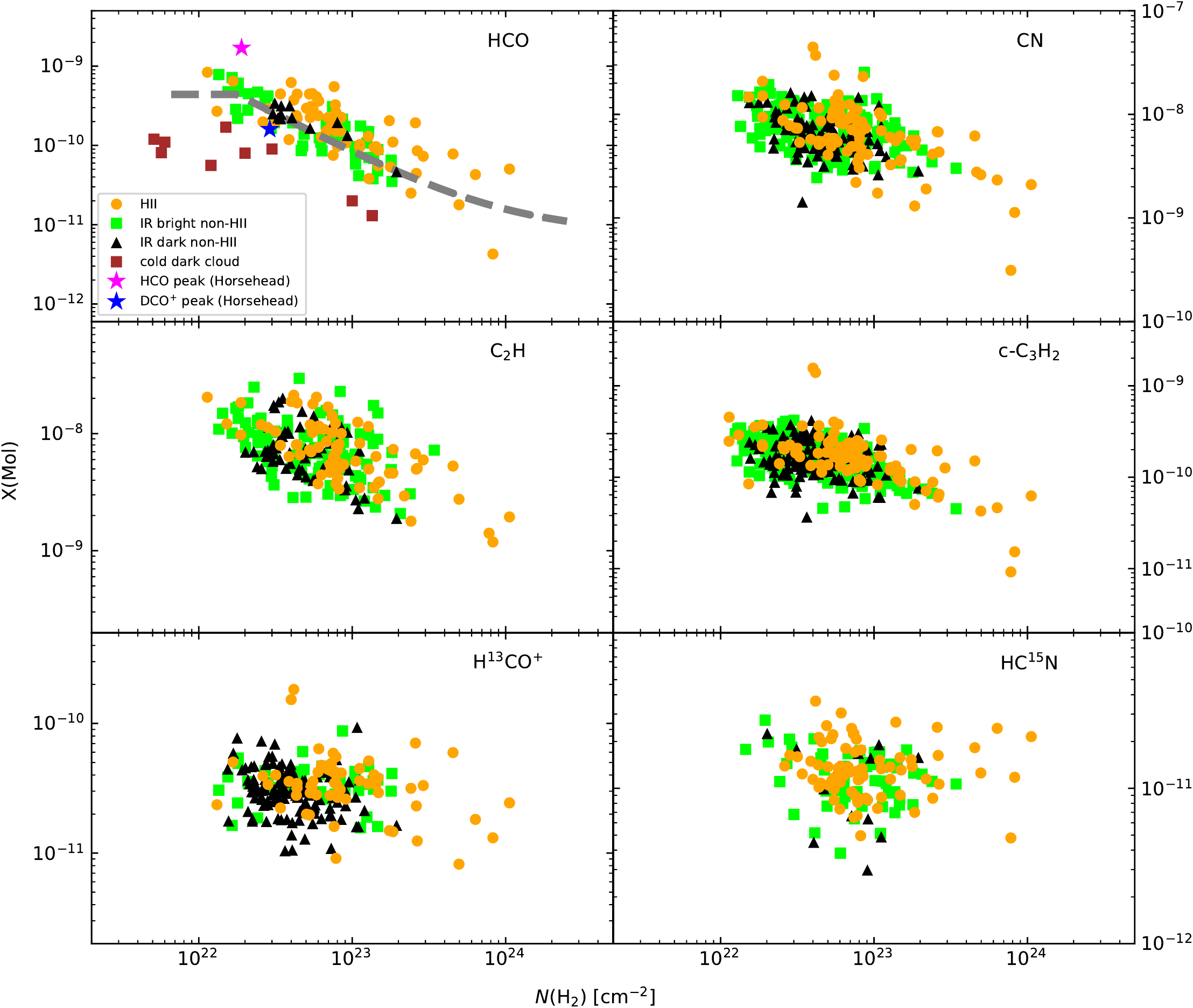}
\caption[]{\label{fig:x_nh2} The fractional abundance of HCO, \hco, \cch, \hccch, \hco\ and \hcnii\  relative to \hh\ as a function of $N$(\hh). The orange indicates \hii\ regions while green and black symbols are IR bright and dark non-\hii\ regions. In the most upper left panel, the brown squares indicate 9 cold dark clouds from \cite{agundez2015}. The purple and blue star symbols are HCO abundances from a PDR and cold gas region, which are resolved by the observation, in the Horsehead nebula \citep{gerin2009}. In the top left panel, the gray dashed line represents a simple abundance jump model.}
\end{figure*}
 
In Fig.\,\ref{fig:x_nh2}, we plot the relationships of $X$(mol) and $N$(\hh) of the observed clumps for the HCO, CN, \cch\ and \hccch\ molecules; these molecules have been found to show good spatial correlations with some PDRs such as the Horsehead nebula, the Orion Bar and Mon R2 \citep{teyssier2004,rizzo2005,gerin2009,ginard2012,cuadrado2015}. In the bottom panels of Fig.\,\ref{fig:x_nh2}, the abundances of dense gas tracers, $X$(\hco) and $X$(\hcnii), are found to be independent of $N$(\hh). On the contrary, $X$(HCO) significantly drops for the clumps with high $N$(\hh) corresponding to a large $A_{V}$. The $X$(CN), $X$(\cch) and $X$(\hccch) display similar trends as seen in the $X$(HCO) plot, but are less pronounced. 

The $X$(HCO) measured in this work may also contain a contribution from HCO molecules originating in cold gas regions of clumps, not only from PDRs. To analyze this quantitatively, we use a simple toy model; it assumes a high abundance ($2.0\times10^{-10}$) of HCO ($A_{V}$ < 5\,mag) in a PDR and a low abundance (8.0$\times10^{-11}$) in the dense molecular region. The resulting average abundance changes as a smooth progression between a PDR and a cold gas region. The gray dashed line on the HCO plot in the most upper left panel of Fig.\,\ref{fig:x_nh2} shows the result of the model. It fits our data points (orange, green and black symbols) and might indicate that the trends of decreasing abundances with $N$(\hh) are caused by abundance jumps from the PDRs into the cold molecular clouds. Besides, at a given $N$(\hh), $X$(HCO) values determined for the \hii\ regions seem to be slightly higher than toward non-\hii\ regions. The HCO molecule is referred to as a good tracer of FUV; its emission in the Horsehead nebula was found in a range $1.5 \lesssim A_{V} \lesssim 3.0$ and its abundance has been found to decrease when the gas is shielded from the FUV radiation \citep{gerin2009}. In contrast, in the Orion Bar HCO emission was found deeper into PDRs toward molecular regions with $5 \lesssim A_{V} \lesssim10$ \citep{schilke2001}. Some of the non-\hii\ region sources (all IR bright non-\hiis\ and potentially some of IR dark non-\hiis) host early stages of star formation (i.e., YSOs) but they do not yet show centimeter free-free emission or mm-RRLs. Thus, their embedded YSOs cannot provide FUV radiation, although shock associated with outflow from them potentially could. However, some of the IR dark non-\hii\ regions are possibly influenced by external radiation fields from nearby star-forming complex regions (e.g., IR dark non-\hii\ regions in M17 SW). Therefore, we presume that bright PDRs are only found in the \hii\ region sources because of abundant UV radiation from their massive stars. This might be the reason that the \hii\ regions show slightly higher $X$(HCO) than the others. We also found that all the ATLASGAL clumps have higher $X$(HCO) compared to nine local cold dark clouds (brown squares) taken from \cite{agundez2015} with distances of around 140$-$500\,pc, at a given $N$(\hh) (the most upper left panel of Fig.\,\ref{fig:x_nh2}). On the contrary, $X$(HCO) (blue star) in the cold gas region in the Horsehead nebula is similar to the abundances in the dust clumps, while $X$(HCO) (purple star) estimated toward the PDR of the Horsehead is higher than any of the sources in the plot. While \cite{gerin2009} measured the HCO abundances from separated cold gas and PDRs, the estimated abundances in this study are likely averages of cold gas and PDRs in the clumps, similar to the result of the simple abundance jump model (gray dashed line). Also, we cannot exclude the possibility that the low detection rates of HCO toward non-\hii\ regions (especially toward IR dark non-\hiis) are due to a lack of UV radiation and/or a distance effect.

\cch\ and \hccch\ are known as tracers of the surface layers of PDRs exposed to a strong UV field (their highest abundances have been found at visual extinctions of 2 mag\,$< A_{V} <$\,5 mag, in typical PDRs; \citealt{rizzo2005,pety2005,ginard2012}). These small hydrocarbons have also shown good spatial correlations with H$_{2}^{*}\,(\nu = 1-0)$ and PAH 8\,\mum\ emission observed toward the Horsehead nebula, the Orion Bar, and Mon R2 \citep{teyssier2004,pety2005,pilleri2013,cuadrado2015}. They were, however, also found in UV shielded regions \citep{beuther2008}. In Fig.\,\ref{fig:x_nh2}, their abundances as well as $X$(CN) decrease with a more moderate slope compared to the $X$(HCO) plot. The abundance decrease of the four molecules might be associated with the transitions of the PDRs and the cold envelopes in the clumps. 

\section{Column density ratios}\label{ch:pdr_mol}

Molecular abundances are used to quantify the number of molecules in dense clumps. However, column density ratios of chemically related molecules are a more direct way to diagnose chemical enhancement of a specific molecule within a clump. Furthermore, these ratios are insensitive to the unknown beam filling factor, the H$_2$ column density and to uncertainties in the assumed physical conditions for molecular species with similar excitation \citep{ginard2012}. We here focus on column density ratios of HCO/\hco\ and \hccch/\cch. Before discussing these column density ratios for each pair, we will briefly explain the proposed formation and destruction of HCO and small hydrocarbons in order to assist in interpreting the column density ratios of these molecules.

\begin{figure}
\centering\includegraphics[width=0.48\textwidth]{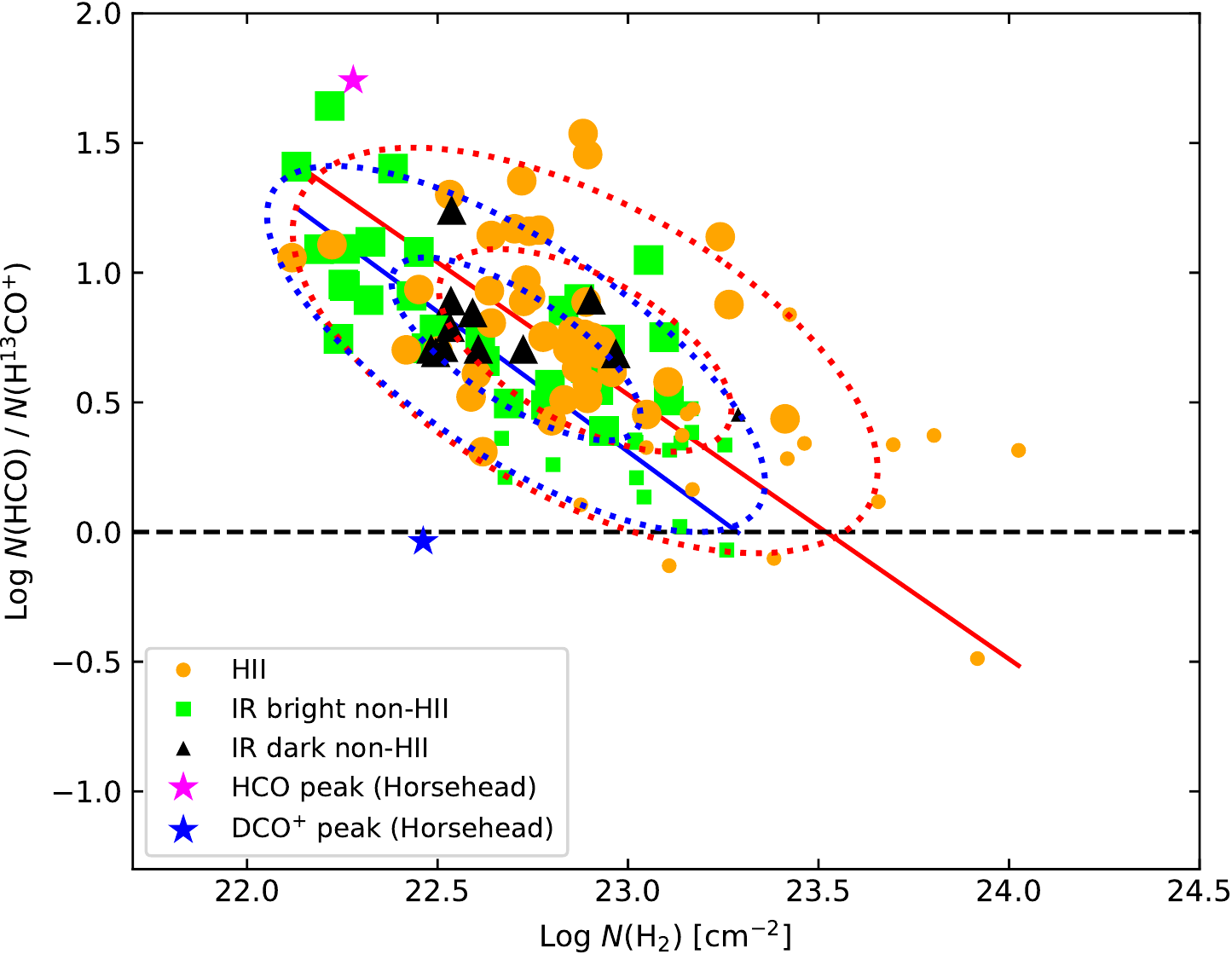}
\includegraphics[width=0.48\textwidth]{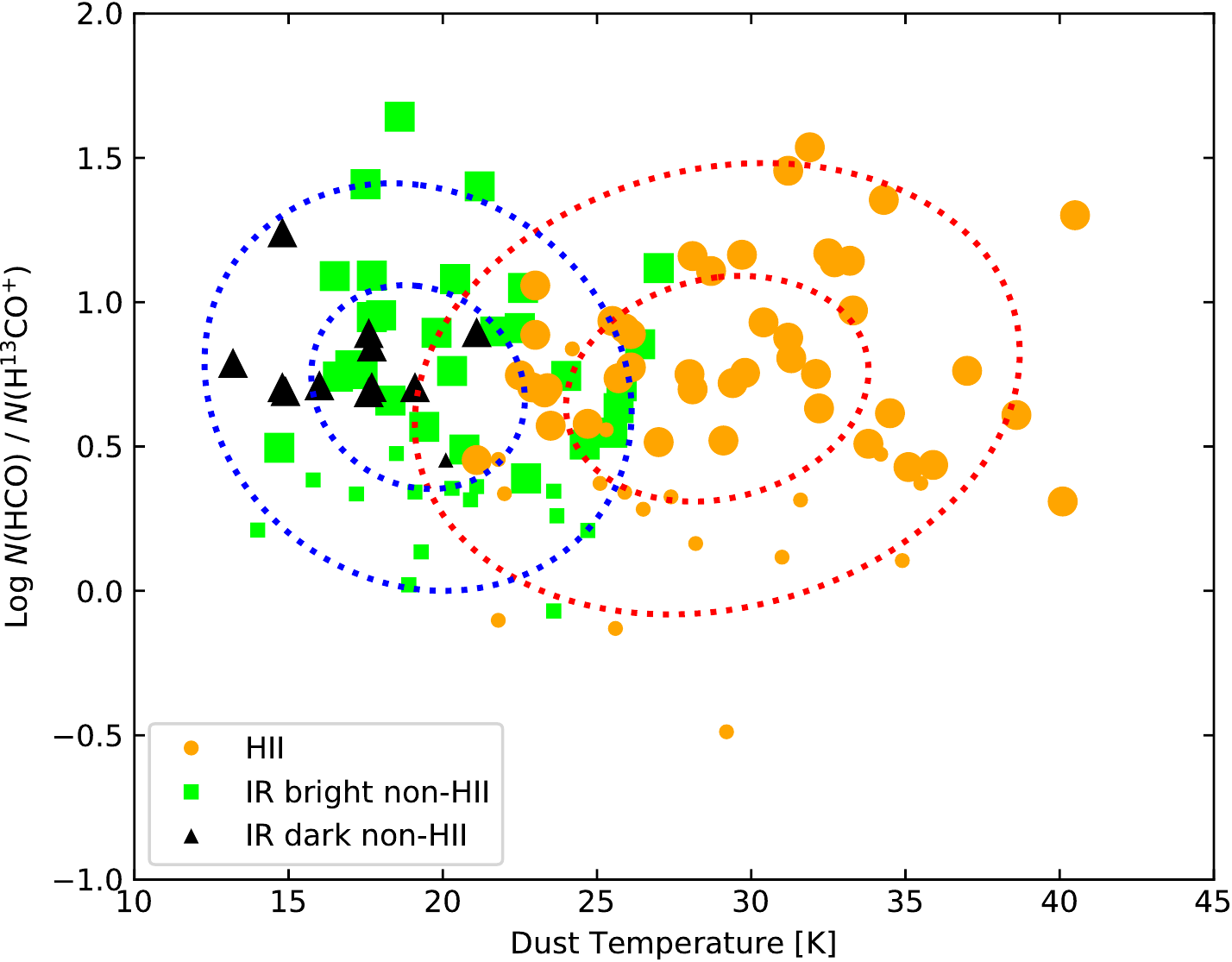}
\caption[]{\label{fig:X_hco_h13cop} Column density ratio of  HCO and \hco\ molecules as a function of $N$(\hh) (upper panel) and $T_{\rm dust}$ (lower panel). The symbols are the same as explained in Fig.\,\ref{fig:x_nh2}. The bigger dots indicate sources with their X(HCO) higher than $10^{-10}$ \citep{gerin2009}. The purple and blue stars (upper panel only) represent $N$(HCO)/$N$(\hco) ratios from the PDR and cold gas region in the Horsehead nebula \citep{gerin2009}. The red and blue dashed line ellipses are confidence ellipses showing distributions of these two populations (\hii\ and non-\hii\ regions) with 1 and 2$\sigma$ levels. In the upper plot, the red and blue lines indicate the best-linear regression fits data points of \hii\ and non-\hii\ regions (IR bright and IR dark non-\hiis). In the upper panel, the black horizontal dashed line presents that $N$(HCO)/$N$(\hco) ratios is 1.}
\end{figure}

\subsection{Formation of HCO and observational HCO/\hco\ column density ratio}

In prior studies \citep{gerin2009,goicoechea2009}, the HCO and \hco\ abundances in cold gas were found to be constant, whereas in PDRs the abundances of these molecules changed due to different chemical reactions. The primary destruction process of \hco\ in PDRs is dissociative recombination with electrons, and this happens quickly: \hco\ $+$ e$^{-}\,\rightarrow\,^{13}$CO $+$ H. On the other hand, the formation of HCO is closely related to FUV radiation in PDRs, and several formation routes have been proposed to describe the high HCO abundance in PDRs via gas-phase reaction or photodissociation \citep{gerin2009}. The gas-phase formation route has two possible chemical reactions. The first one is a vital formation route of HCO in FUV shielded regions \citep{schenewerk1988,gerin2009}: 
\begin{equation}\label{eq:1st_route}
\noindent {\rm Metals~(Mg~or~Fe)} + {\rm HCO^{+}}
\rightarrow {\rm HCO} + {\rm metals^{+}}~{\rm (Mg^{+}~or~Fe^{+})}. 
\end{equation}
The most plausible pure gas-phase route is a reaction of atomic oxygen with carbon radicals in PDRs \citep{watt1983, leung1984,schenewerk1988}:
\begin{equation}\label{eq:3rd_route}
\noindent {\rm O} + {\rm CH_{2}} \rightarrow {\rm HCO} + {\rm H}. 
\end{equation}
Another suggested formation route is the gas-grain reaction through FUV radiation: photodissociation of formaldehyde (H$_{2}$CO) or grain photodesorption. The first route by FUV has been proposed by \cite{schilke2001},
\begin{equation}\label{eq:2nd_route}
\noindent {\rm H_{2}CO} + {\rm photon} \rightarrow {\rm HCO} + {\rm H}. 
\end{equation}
The second route is by grain photodesorption: HCO forms on grain mantles, and subsequently it is desorpted from the grains into the gas-phase via thermal or/and photo-desorption processes \citep{gerin2009}. In cold gas regions below $\sim$30\,K, thermal desorption does not play a primary role \citep{gerin2009}, whereas, in warm regions like \hii\ regions, it possibly contributes to HCO desorption. If thermal desorption is not essential, ice-mantle photo-desorption induced by FUV radiation is an alternative process \citep{willacy1993,bergin1995}. The fact that high $X$(HCO) is always found in PDRs also supports this photo-desorption process \citep{schenewerk1988,schilke2001, gerin2009}.

Since HCO and HCO$^+$ result from different chemical reactions in PDRs and cold gas regions, we compare column densities of these molecules to constrain the origin of the detected HCO and \hco\ toward ATLASGAL clumps. \hco\ is used to avoid high optical thicknesses. In the upper panel of Fig.\,\ref{fig:X_hco_h13cop}, $N$(HCO)/$N$(\hco) ratios decrease with an increase of $N$(\hh), which is closely related to a decrease of $X$(HCO) because $X$(\hco) is mostly independent of $N$(\hh) (Fig.\,\ref{fig:x_nh2}). Besides, results of linear fits for different source types show that \hii\ region sources (red line, slope = $-$1.02) are slightly shifted toward higher $N$(HCO)/$N$(\hco) ratios than non-\hii\ region sources (blue line, slope = $-$1.08). Visually, the separation (as their confidence ellipses) between them is small, and the slope of their fitting lines are similar. However, a small $p$-value (0.01) from a 2-dimensional KS test on the two group distributions indicates that these groups are significantly different samples.

We added $X$(HCO), and $N$(HCO)/$N$(\hco) values observed toward a PDR and a cold gas region in the Horsehead nebula \citep{gerin2009} in the upper panel of Fig\,\ref{fig:X_hco_h13cop}, as a reference for the origin of HCO toward ATLASGAL clumps. 
$N$(HCO)/$N$(\hco) ratios toward \hii\ (orange color) and non-\hii\ regions (green and black colors) are distributed between the values in the PDR (purple star, HCO peak in the Horsehead nebula) and the cold gas region (blue star, DCO$^{+}$ peak in the Horse nebula). The HCO in the cold gas area is considered to be originated from the low column density surface of the cold gas regions, not from the UV shielded cold gas region.    
\begin{figure}
\centering
\includegraphics[width=0.49\textwidth]{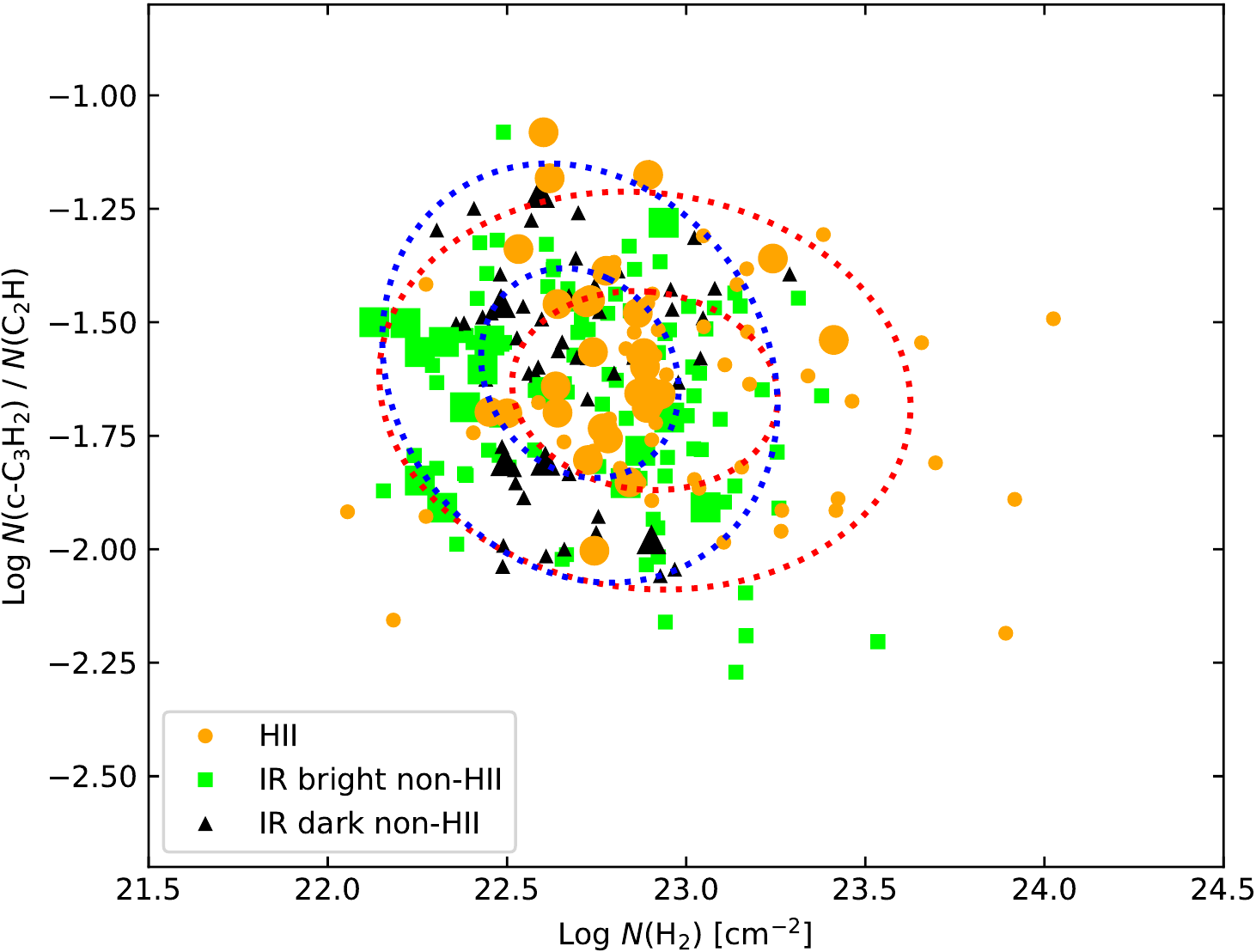}
\includegraphics[width=0.48\textwidth]{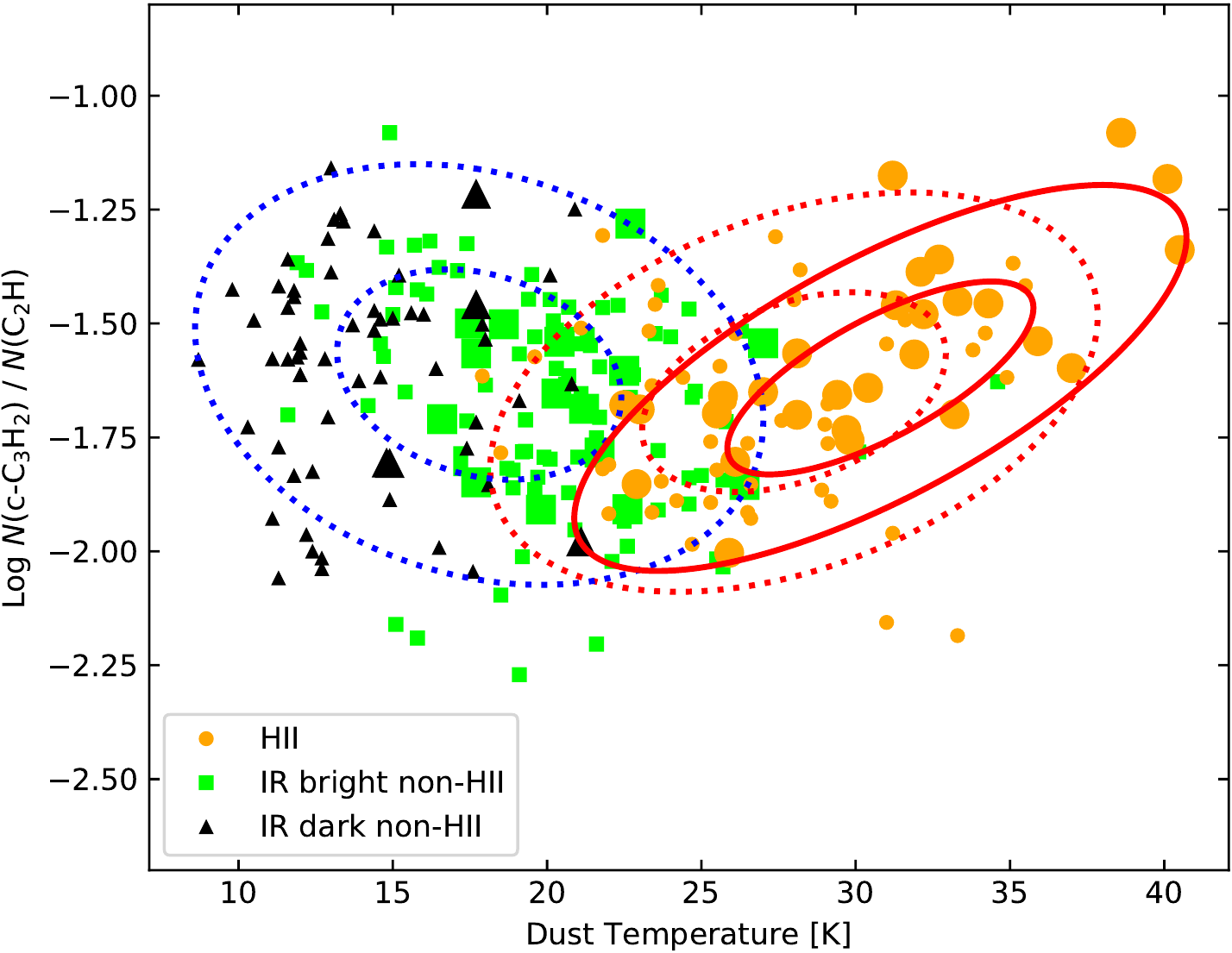}
\caption[]{\label{fig:X_cch_c3h2} $Upper:$ Column density ratio of \hccch\ and \cch\ as a function of $N$(\hh). $Lower:$ $N$(\hccch)/$N$(\cch) ratios as a function of $T_{\rm clump}$. Same symbols and lines as Fig.\,\ref{fig:X_hco_h13cop}. }
\end{figure}

The clumps with bigger symbol size indicate $X$(HCO) greater than 10$^{-10}$, which \cite{gerin2009} proposed as an indicator of the presence of FUV radiation. All of them are located above a black dashed line corresponding to $N$(HCO)/$N$(\hco) of 1. According to modelling results for the Horsehead nebula \citep{gerin2009}, $X$(HCO) $\gtrsim$ 10$^{-10}$ and $N$(HCO)/$N$(\hco) $\gtrsim$ 1 together suggest on-going FUV photochemistry. In addition to that, the observational and modeling results for Mon R2 \citep{ginard2012} also found HCO abundant only in PDRs with FUV radiation field ($G_{0}$ $< 10^{3}$ in units of the Habing field), but not with HUV (like the center of Mon R2, $5\times10^{5}$). A high HCO/HCO$^{+}$ ratio (> 0.2) was only found in the low UV irradiated region. Therefore, the HCO abundances for the clumps with $X$(HCO) $\gtrsim$ 10$^{-10}$ and $N$(HCO)/$N$(\hco) $\gtrsim$ 1 probably probe the presence of PDR illuminated by FUV radiation in clumps. It is likely one of the reasons why high $X$(HCO) and $N$(HCO)/$N$(\hco) ratios are not always found for \hii\ region sources but also for non-\hii\ regions (especially IR dark non-\hii\ regions marked with black symbols). In previous HCO observations toward starless cores (e.g., \citealt{ frau2012,agundez2015}), HCO lines were often detected toward sources with \hh\ column densities lower than $10^{23}$\,cm$^{-2}$ ($\sim$ some $10^{22}$\,cm$^{-2}$). High $X$(HCO) ($\gtrsim 10^{-10}$) in our sources is found mostly from the clumps with $N$(\hh) $< 10^{23}$\,cm$^{-2}$. HCO emission in the clumps with \hh\ column density of $\geq 10^{23}$\,cm$^{-2}$ is detectable because bright, massive star-forming regions provide enough radiation to irradiate such high \hh\ column density regions, or their surfaces are illuminated by nearby massive stars. Also, it is possible that the UV radiation from the deeply embedded high-mass stars can still penetrate the dense clumps and create the embedded PDRs. The line widths of HCO emission found toward the \hii\ region clumps are significantly broader than those of \hco, which is abundant in cold, dense gas regions as seen in the Horsehead nebula \citep{gerin2009}, with a small fitting uncertainty (see Fig.\,\ref{fig:linewdith_all}). This also supports that the HCO emission lines detected toward the clumps with high \hh\ column densities are not emitted from the dense regions. 

In the lower panel of Fig.\,\ref{fig:X_hco_h13cop}, we compare the column density ratio with dust temperature because one possible HCO formation route is related to thermal grain desorption. The plot shows that thermal-desorption from grains does not seem to play an important role since no correlation is found between the ratios and dust temperatures. The lack of correlation with the dust temperature might indirectly explain the reason that we also found high $X$(HCO) for non-\hii\ regions, not only \hii\ regions. Even if grain thermal-desorption processes contributes to the observed abundances, it is difficult to probe such reaction with our current low angular resolution observational data that average HCO emission and dust temperatures over large clumps within single-dish telescope beams. Therefore, the possibility of thermal desorption is not fully excluded from the formation of HCO.
If HCO is mainly formed on grain mantles and is desorpted by photons, we need to compare HCO with H$_{2}$CO, CH$_{3}$O and CH$_{3}$OH. These molecules are also formed on grain mantles by hydrogenation reactions of CO-ice \citep{tielens1997,charnley1997}. Studying both HCO and the other molecules allows us to test whether grain photo-desorption processes cause the main formation reaction of HCO in massive clumps. In addition, we also need to have better angular resolution to separate relative contributions of HCO emission from PDRs and cold gas regions. Our observed abundances are lower limit because their emission is averaged over the clumps. Observing with interferometers toward the HCO clumps will allow us more accurate HCO abundances and to have a better understanding of its formation and destruction, and furthermore, complex organic molecules that are formed after HCO. 

\subsection{Formation of small hydrocarbons and observational \mbox{\hccch/\cch} column density ratio}

The formation of small hydrocarbons is still not fully understood, but several processes of their formation in PDRs have been proposed: gas-phase reactions and grain-surface reactions such as photodestruction of PAHs or very small grains (VSGs). In highly UV-illuminated PDRs with high gas temperatures (a few hundreds  $\sim$ 1000\,K), within an atomic layer of the PDRs, the gas-phase formation of small hydrocarbon dominates when ionized carbon and electrons are abundant \citep{cuadrado2015}. Under these conditions, \cch\ is formed before \hccch\ because \hccch\ needs an additional carbon atom. 
\cch\ can be formed by recombination of ${\rm C_{2}H^{+}}$, ${\rm C_{2}H_{2}^{+}}$, or ${\rm C_{2}H_{3}^{+}}$ with electrons in PDRs (e.g., \citealt{mookerjea2012,cuadrado2015}) via barrierless hydrogenation reactions, for example, 
\begin{equation}\label{eq:cch_route1}
{\rm C_{2}H_{3}^{+}} + e^{-} \rightarrow {\rm C_{2}H~and~2 H}.
\end{equation}
Also, \cch\ can be formed by photodissociation of acetylene (C$_{2}$H$_{2}$) in gas phase \citep{lee1984};
\begin{equation}\label{eq:cch_route2}
    {\rm C_{2}H_{2}} + {\rm h}\nu \rightarrow {\rm C_{2} H} + {\rm H}. 
\end{equation}

For the gas-phase formation of \hccch, C$_{2}$H$_{2}$ and C$^{+}$ form C$_{3}$H$^{+}$ that reacts with H$_{2}$ and then produces the linear and cyclic C$_{3}$H$_{3}^{+}$ isomers \citep{maluendes1993,mcewan1999}. Through dissociative recombination of these molecules \citep{fosse2001}, linear and cyclic-C$_{3}$H$_{2}$ are formed as
\begin{equation}\label{eq:c3h2_route}
c/l{\rm -C_{3}H_{3}^{+}} + e^{-} \rightarrow c/l{\rm -C_{3}H_{2}} + {\rm H}. 
\end{equation}

Both C$_3$H$_2$ isomers are destroyed by photodissociation in a strong-UV radiation field, which in general should destroys small hydrocarbons. Thus, their observed abundances in such PDRs cannot be explained with gas-phase reactions only. Besides, the \hccch\ abundances in Mon R2 were found to be higher close to the UV-exposed PDR than the \cch\ abundances \citep{pilleri2013}. The formation sequences mentioned above cannot easily result in the high \hccch\ abundances, and reproducing observed \hccch\ values with PDR models has been difficult (e.g., \hccch\ abundances in M8, \citealt{tiwari2019}). Another possible way to produce \hccch\ is grain-surface formation via photons. The photodissociation of small PAHs or VSGs in PDRs provides fresh small hydrocarbons in highly-illuminated PDRs \citep{fuente2003,pety2005} and laboratory experiments also have demonstrated the production of small hydrocarbons from small PAHs (number of carbons $\leq$ 24, \citealt{useli_bacchitta2007}). In particular, PAH-related photochemistry enhances the abundance of \hccch\ in PDRs via dissociative recombination of C$_{3}$H$_{3}^{+}$ with electrons ejected from PAHs or fragmentation of PAHs \citep{mookerjea2012,pilleri2013}.

Although those small hydrocarbons are well-known PDR tracers, their abundances vary from one PDR to the other (e.g., \citealt{pety2005, mookerjea2012, pilleri2013, cuadrado2015, tiwari2019}). By investigating their abundances for ATLASGAL clumps, we may gain insight into the origin of the detected small hydrocarbons (i.e., \cch\ and \hccch). Unlike the other PDR tracer, HCO, $N$(\hccch)/$N$(\cch) in the upper panel of Fig.\,\ref{fig:X_cch_c3h2} does not show any correlation with $N$(\hh). The presence of UV radiation and C$^{+}$ is a necessary factor for the formation of \cch\ and \hccch, and thus, we mark with bigger dots sources with $X$(HCO) $\gtrsim 10^{-10}$ and $N$(HCO)/$N$(\hco) $\gtrsim$ 1, which are proposed as diagnostics of the presence of FUV radiation fields and probably C$^+$ \citep{gerin2009}. However, these sources also do not show any correlation with any of the parameters. 

While HCO is found toward a few sources, these small hydrocarbons are ubiquitous in molecular clouds although they are probably more abundant in PDRs. To compare with our beam ($\sim 30''$ corresponding to a size of 0.73\,pc at a median distance of 5\,kpc for the observed sources), highly illuminated or dense PDRs surrounding young \hii\ regions are spatially very small and thin \citep[$\sim$ 0.001\,pc,][]{pilleri2013}. Therefore, the contribution of small hydrocarbons in the cold and low-density molecular envelope surrounding such PDRs and \hii\ regions is presumably dominating the observed emission, as in the case of the molecular envelope of Mon R2. The small hydrocarbons column densities we find in this study are close to averaged values of the molecular envelope and PDRs in Mon R2 or to those of the envelope \citep{pilleri2013}. We suggest that the origin of the measured abundances of \cch\ and \hccch\ might be from the envelope of clumps rather than the PDR regions.

\begin{figure}
\centering
\includegraphics[width=0.49\textwidth]{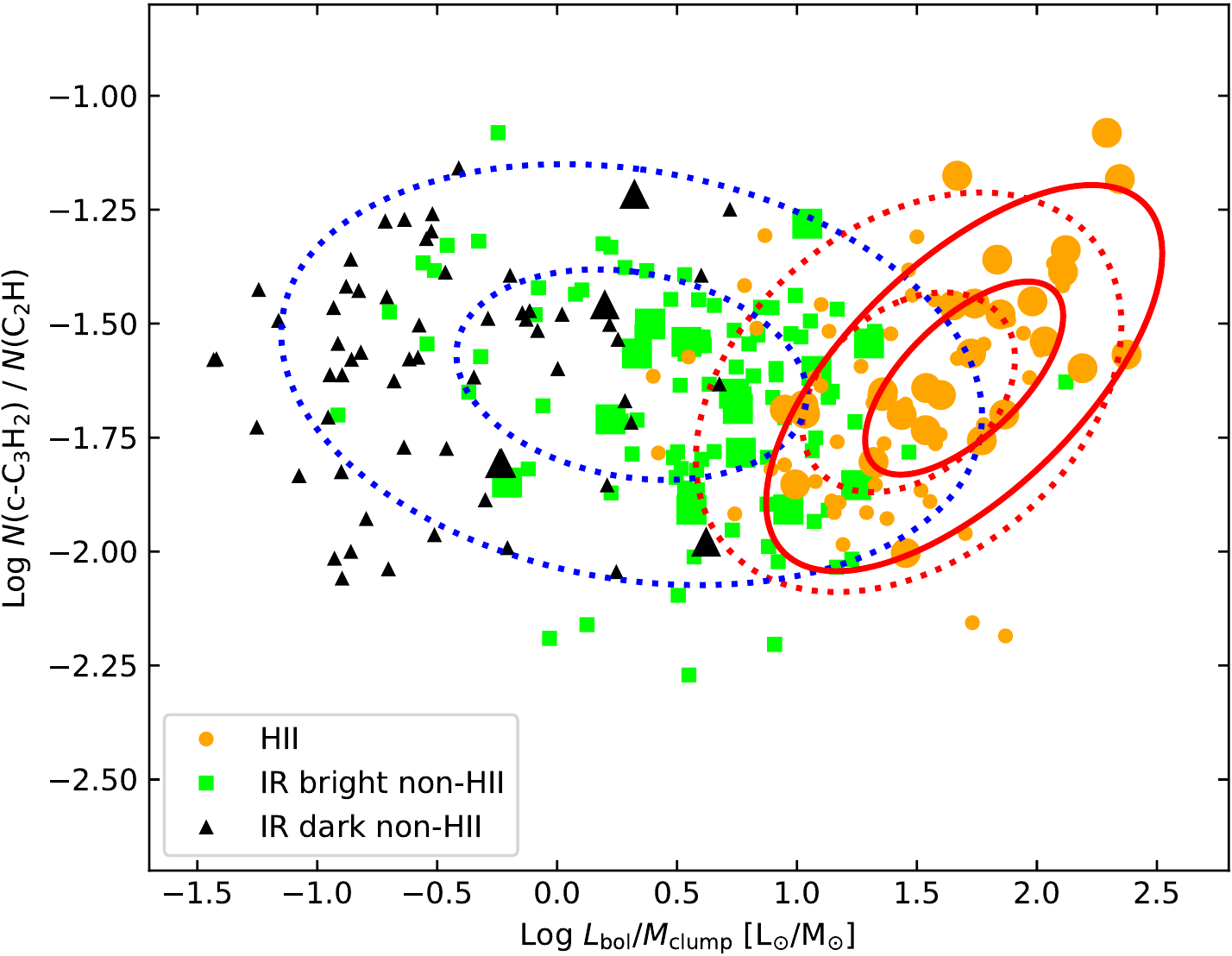}
\caption[]{\label{fig:X_cch_c3h2_rlumass} Column density ratio of \hccch\ and \cch\ as a function of clump bolometirc luminosity over their mass ($L_{\rm bol}/M_{\rm clump}$). Same symbols and lines as in Fig.\,\ref{fig:X_hco_h13cop}. }
\end{figure}
In low-density ($n_{\rm H_{2}} \sim 5\times10^{5}$\,cm$^{-3}$) envelopes with low kinetic temperature ($T_{\rm kin}\sim35$\,K), many species are presumably locked in the icy mantles of dust grains, and some fraction of them sublimate into the gas-phase during the collapse of prestellar cores \citep{pilleri2013}. We, therefore, investigated the influences of the kinetic temperature on the small hydrocarbon abundances in ATLASGAL clumps. The lower panel of Fig.\,\ref{fig:X_cch_c3h2} shows the $N$(\hccch)/$N$(\cch) ratio versus the dust temperature of the clumps. Without separating different source groups (\hii\ and non-\hii\ regions), the ratios seem to be constant over the range of dust temperatures (10$-$40\,K) with a large scatter. Previously, \cite{viti2004} and \cite{pilleri2013} showed that final abundances of these small hydrocarbons did not show differences in the temperature range of 10$-$35\,K because most of the molecular species are still on grain mantles below $T_{\rm dust}=35$\,K. However, when we focus on only \hii\ regions (orange dots), there is a trend of $N$(\hccch)/$N$(\cch) ratios increasing with dust temperature, and in particular, the sources with high HCO abundances (bigger dots) show an even better correlation (red solid-line confidence ellipses). The dust temperatures of clumps are associated with the evolution of the clumps. In Fig.\,\ref{fig:X_cch_c3h2_rlumass}, we compare the small hydrocarbons with bolometric luminosity over clump mass ($L_{\rm bol}/M_{\rm clump}$) that is used as an evolutionary stage indicator of clumps, and thus higher $L_{\rm bol}/M_{\rm clump}$ refer to more evolved clumps. It is evident that high $N$(\hccch)/$N$(\cch) ratios are found toward more evolved clumps that are associated with \hii\ regions and have high HCO abundances indirectly probing the presence of FUV radiation fields. According to results of the models for the molecular envelope in Mon R2 and DR21 \citep{mookerjea2012,pilleri2013}, in the cold and low-density molecular envelopes of the regions, grain-surface processes (i.e., CO freeze-out) and time-dependent effects become an important role in the formation and destruction of these molecules. In particular, while abundances of \cch\ and \hccch\ do not vary until $10^{4}$ years, after about $10^{5}$ years with the development of \uchii\ regions, the $N$(\hccch)/$N$(\cch) ratios steeply go up. Depending on early freeze-out fractions, which control the initial small hydrocarbons abundances in the gas-phase, final abundances of these molecules predicted in the models vary. Since we do not have any information about the initial physical conditions in our clumps, it is hard to predict how the gain-surface processes and time-dependent effects affect our observed results. The Stratospheric Observatory for Infrared Astronomy (SOFIA) observations toward M17SW \citep{perez-beaupuits2012} showed that [C {\sc{ii}}] emission was found interestingly throughout deeper parts of the molecular cloud, and a significant fraction of its emission was not associated with other PDR tracers (i.e., CO $J = 13-12$ and [C {\sc{i}}]). This means that C$^{+}$ gas, probed by [C {\sc{ii}}] spectral lines, can exist in deeper molecular clouds. Thus, we cannot exclude that C$^{+}$ can contribute to the formation of small hydrocarbon via gas-phase reactions in such molecular envelop regions. In addition, photodissociation of PAH molecules in different strengths of UV radiation fields can add some fraction of small hydrocarbons abundances in gas-phase. According to \cite{murga2020}, in high-UV illuminated regions (i.e., the Orion Bar), the PAH dissociation becomes more important at $A_{\rm V} > 3.5$ molecular gas region than gas-phase reaction to produce C$_{2}$H$_{2}$, which is one of precursors of \cch\ and \hccch\ as mentioned above. On the other hand, in low-UV radiation, pure gas-phase reaction is a main route to increase the production rate of C$_{2}$H$_{2}$. Since the more evolved clumps shown in Fig.\,\ref{fig:X_cch_c3h2_rlumass} emit stronger UV radiation, it is possible that the measured abundances of the small hydrocarbons are still associated with the photodissociation. We suggest that the small hydrocarbons we observe toward clumps hosting \hii\ regions likely originate mostly in the molecular envelope around dense PDRs and \hii\ regions. In particular, abundances of these molecules might be a combined result of gain-surface processes controlling initial molecular abundances in the gas-phase and subsequent time-dependent evolution in environments with a significant C$^{+}$ abundance. Also, some fraction of these small hydrocarbon abundances in the \hii\ sources can be resulted of photodissociation of PAH molecules by the UV radiation from the newly born high-mass stars. 

\section{Summary and conclusion}\label{ch:summary}

We have investigated 10 molecular transitions from HCO, \hco, HOC$^{+}$, \cch, \hccch, CN, \hcni, \hcnii, \hnc\ and \co\ covered in an unbiased molecular line survey of the 3\,mm band that was observed using the IRAM 30\,m telescope toward 409 ATLASGAL clumps. The ATLASGAL clumps were divided into three groups based on the presence of \hii\ regions and infrared emission: \hii, IR bright non-\hii, and dark non-\hii\ region sources. We carried out an analysis of the column density and abundance of the selected molecules, and our main results are as follows:
\begin{itemize}
    \item [$-$] \co, \hco, \cch, \hccch, CN, and \hnc. \hcni\ show high detection rates (higher than 94\,\%), whereas \hcni, HCO and \hcnii\ are detected with detection rates of 75\%, 32\%, and 39\%, respectively. The non-detections of the HCO, \hcni\ and \hcnii\ transitions are mostly from the IR-bright and IR-dark non-\hii\ regions, while high detection rates of the molecular transitions toward \hii\ region sources were found, namely HCO in 66\%, \hcni\ in 93\% and \hcnii\ in 79\% of the regions. 
    
    \item[$-$] While the abundances of high column density tracers (i.e., \hco\, and \hcnii) are almost constant over the range of \hh\ column densities, the abundances of HCO, CN, \cch\ and \hccch\ drop with an increase of \hh\ column density. In particular, the HCO abundances are prominently reduced in high \hh\ column density, and they seem higher toward \hii\ regions than toward non-\hii\ regions for a given \hh\ column density. 
    
    \item[$-$]We also find that $N$(HCO)/$N$(\hco) ratios decrease as \hh\ column density increase, and 61 clumps have $X$(HCO) $\gtrsim 10^{-10}$ and $N$(HCO)/$N$(\hco) $\gtrsim 1$. This implies that the HCO detected toward ATLASGAL clumps is likely connected to PDRs, and the sources with high HCO abundances are associated with on-going FUV chemistry in their PDRs.  However, due to low angular resolution of our data, the measured HCO abundances in the ATLASGAL clumps are averages over their PDRs and FUV shielded molecular gas regions. 
    
    \item[$-$]The \hccch/\cch\ ratios toward the dust clumps are constant with \hh\ column density and dust temperature, although with large scatter. However, toward only \hii\ regions having high HCO abundances, their \hccch/\cch\ ratios evidently rise with dust temperatures. Especially, high  \hccch/\cch\ ratios are found toward more evolved clumps with high $L_{\rm bol}$/$M_{\rm clump}$. These results show similar trends with the model results of \cite{pilleri2013} that predict increasing \hccch/\cch\ ratios with the time after the age of $10^{5}$ years. Therefore, the measured abundances of small hydrocarbons in this study are possibly results of gain-surface processes (CO freeze-out fractions) and time-dependent effects in the clumps rather than in the PDRs. In addition, some fraction of the measured abundances toward the \hii\ sources can be added as the result of the photodissociation of PAH molecules. 

\end{itemize}{}

\section*{Acknowledgements}

\addcontentsline{toc}{section}{Acknowledgements} We would like to thank the referee for their constructive
comments and suggestions that have helped to improve this paper. Won-Ju\, Kim was supported for this research through a stipend from the International Max Planck Research School (IMPRS) for Astronomy and Astrophysics at the Universities of Bonn and Cologne. This work was partially funded by the Collaborate Research Council 956, subproject A6, funded by the Deutsche Forschungsgemeinschaft. The ATLASGAL project is a collaboration between the Max-Planck-Gesellschaft, the European Southern Observatory (ESO) and the Universidad de Chile. It includes projects E-181.C-0885, E-078.F-9040(A), M-079.C-9501(A), M-081.C-9501(A) plus Chilean data. This document was produced using the Overleaf web application, which can be found at www.overleaf.com.

\bibliographystyle{aa}
\bibliography{mspdr_wjkim.bbl}

\begin{thebibliography}{74}
\expandafter\ifx\csname natexlab\endcsname\relax\def\natexlab#1{#1}\fi

\bibitem[{{Ag{\'u}ndez} {et~al.}(2015){Ag{\'u}ndez}, {Cernicharo}, \&
  {Gu{\'e}lin}}]{agundez2015}
{Ag{\'u}ndez}, M., {Cernicharo}, J., \& {Gu{\'e}lin}, M. 2015, \aap, 577, L5

\bibitem[{{Bacmann} {et~al.}(2002){Bacmann}, {Lefloch}, {Ceccarelli},
  {Castets}, {Steinacker}, \& {Loinard}}]{bacmann2002}
{Bacmann}, A., {Lefloch}, B., {Ceccarelli}, C., {et~al.} 2002, \aap, 389, L6

\bibitem[{{Bergin} {et~al.}(1995){Bergin}, {Langer}, \&
  {Goldsmith}}]{bergin1995}
{Bergin}, E.~A., {Langer}, W.~D., \& {Goldsmith}, P.~F. 1995, \apj, 441, 222

\bibitem[{{Beuther} {et~al.}(2008){Beuther}, {Semenov}, {Henning}, \&
  {Linz}}]{beuther2008}
{Beuther}, H., {Semenov}, D., {Henning}, T., \& {Linz}, H. 2008, \apjl, 675,
  L33

\bibitem[{{Boger} \& {Sternberg}(2005)}]{boger2005}
{Boger}, G.~I. \& {Sternberg}, A. 2005, \apj, 632, 302

\bibitem[{{Caselli} {et~al.}(1999){Caselli}, {Walmsley}, {Tafalla}, {Dore}, \&
  {Myers}}]{caselli1999}
{Caselli}, P., {Walmsley}, C.~M., {Tafalla}, M., {Dore}, L., \& {Myers}, P.~C.
  1999, \apjl, 523, L165

\bibitem[{{Charnley} {et~al.}(1997){Charnley}, {Tielens}, \&
  {Rodgers}}]{charnley1997}
{Charnley}, S.~B., {Tielens}, A.~G.~G.~M., \& {Rodgers}, S.~D. 1997, \apjl,
  482, L203

\bibitem[{{Contreras} {et~al.}(2013){Contreras}, {Schuller}, {Urquhart},
  {Csengeri}, {Wyrowski}, {Beuther}, {Bontemps}, {Bronfman}, {Henning},
  {Menten}, {Schilke}, {Walmsley}, {Wienen}, {Tackenberg}, \&
  {Linz}}]{contreras2013}
{Contreras}, Y., {Schuller}, F., {Urquhart}, J.~S., {et~al.} 2013, \aap, 549,
  A45

\bibitem[{{Csengeri} {et~al.}(2016){Csengeri}, {Leurini}, {Wyrowski},
  {Urquhart}, {Menten}, {Walmsley}, {Bontemps}, {Wienen}, {Beuther}, {Motte},
  {Nguyen-Luong}, {Schilke}, {Schuller}, {Zavagno}, \& {Sanna}}]{csengeri2016b}
{Csengeri}, T., {Leurini}, S., {Wyrowski}, F., {et~al.} 2016, \aap, 586, A149

\bibitem[{{Cuadrado} {et~al.}(2015){Cuadrado}, {Goicoechea}, {Pilleri},
  {Cernicharo}, {Fuente}, \& {Joblin}}]{cuadrado2015}
{Cuadrado}, S., {Goicoechea}, J.~R., {Pilleri}, P., {et~al.} 2015, \aap, 575,
  A82

\bibitem[{{Foss{\'e}} {et~al.}(2001){Foss{\'e}}, {Cernicharo}, {Gerin}, \&
  {Cox}}]{fosse2001}
{Foss{\'e}}, D., {Cernicharo}, J., {Gerin}, M., \& {Cox}, P. 2001, \apj, 552,
  168

\bibitem[{{Frau} {et~al.}(2012){Frau}, {Girart}, \& {Beltr{\'a}n}}]{frau2012}
{Frau}, P., {Girart}, J.~M., \& {Beltr{\'a}n}, M.~T. 2012, \aap, 537, L9

\bibitem[{{Fuente} {et~al.}(2005){Fuente}, {Garc{\'{\i}}a-Burillo}, {Gerin},
  {Teyssier}, {Usero}, {Rizzo}, \& {de Vicente}}]{fuente2005}
{Fuente}, A., {Garc{\'{\i}}a-Burillo}, S., {Gerin}, M., {et~al.} 2005, \apjl,
  619, L155

\bibitem[{{Fuente} {et~al.}(1993){Fuente}, {Martin-Pintado}, {Cernicharo}, \&
  {Bachiller}}]{fuente1993}
{Fuente}, A., {Martin-Pintado}, J., {Cernicharo}, J., \& {Bachiller}, R. 1993,
  \aap, 276, 473

\bibitem[{{Fuente} {et~al.}(2003){Fuente}, {Rodr{\i}guez-Franco},
  {Garc{\i}a-Burillo}, {Mart{\i}n-Pintado}, \& {Black}}]{fuente2003}
{Fuente}, A., {Rodr{\i}guez-Franco}, A., {Garc{\i}a-Burillo}, S.,
  {Mart{\i}n-Pintado}, J., \& {Black}, J.~H. 2003, \aap, 406, 899

\bibitem[{{Garc{\'{\i}}a-Burillo} {et~al.}(2002){Garc{\'{\i}}a-Burillo},
  {Mart{\'{\i}}n-Pintado}, {Fuente}, {Usero}, \& {Neri}}]{garcia-burillo2002}
{Garc{\'{\i}}a-Burillo}, S., {Mart{\'{\i}}n-Pintado}, J., {Fuente}, A.,
  {Usero}, A., \& {Neri}, R. 2002, \apjl, 575, L55

\bibitem[{{Gerin} {et~al.}(2009){Gerin}, {Goicoechea}, {Pety}, \&
  {Hily-Blant}}]{gerin2009}
{Gerin}, M., {Goicoechea}, J.~R., {Pety}, J., \& {Hily-Blant}, P. 2009, \aap,
  494, 977

\bibitem[{{Gerner} {et~al.}(2014){Gerner}, {Beuther}, {Semenov}, {Linz},
  {Vasyunina}, {Bihr}, {Shirley}, \& {Henning}}]{gerner2014}
{Gerner}, T., {Beuther}, H., {Semenov}, D., {et~al.} 2014, \aap, 563, A97

\bibitem[{{Giannetti} {et~al.}(2014){Giannetti}, {Wyrowski}, {Brand},
  {Csengeri}, {Fontani}, {Walmsley}, {Nguyen Luong}, {Beuther}, {Schuller},
  {G{\"u}sten}, \& {Menten}}]{giannetti2014}
{Giannetti}, A., {Wyrowski}, F., {Brand}, J., {et~al.} 2014, \aap, 570, A65

\bibitem[{{Ginard} {et~al.}(2012){Ginard}, {Gonz{\'a}lez-Garc{\'{\i}}a},
  {Fuente}, {Cernicharo}, {Alonso-Albi}, {Pilleri}, {Gerin},
  {Garc{\'{\i}}a-Burillo}, {Ossenkopf}, {Rizzo}, {Kramer}, {Goicoechea},
  {Pety}, {Bern{\'e}}, \& {Joblin}}]{ginard2012}
{Ginard}, D., {Gonz{\'a}lez-Garc{\'{\i}}a}, M., {Fuente}, A., {et~al.} 2012,
  \aap, 543, A27

\bibitem[{{Goicoechea} {et~al.}(2009){Goicoechea}, {Pety}, {Gerin},
  {Hily-Blant}, \& {Le Bourlot}}]{goicoechea2009}
{Goicoechea}, J.~R., {Pety}, J., {Gerin}, M., {Hily-Blant}, P., \& {Le
  Bourlot}, J. 2009, \aap, 498, 771

\bibitem[{{Goldsmith} {et~al.}(1986){Goldsmith}, {Irvine}, {Hjalmarson}, \&
  {Ellder}}]{goldsmith1986}
{Goldsmith}, P.~F., {Irvine}, W.~M., {Hjalmarson}, A., \& {Ellder}, J. 1986,
  \apj, 310, 383

\bibitem[{{Hogerheijde} {et~al.}(1995){Hogerheijde}, {Jansen}, \& {van
  Dishoeck}}]{hogerheijde1995}
{Hogerheijde}, M.~R., {Jansen}, D.~J., \& {van Dishoeck}, E.~F. 1995, \aap,
  294, 792

\bibitem[{{Jansen} {et~al.}(1995){Jansen}, {Spaans}, {Hogerheijde}, \& {van
  Dishoeck}}]{jansen1995}
{Jansen}, D.~J., {Spaans}, M., {Hogerheijde}, M.~R., \& {van Dishoeck}, E.~F.
  1995, \aap, 303, 541

\bibitem[{{Jin} {et~al.}(2015){Jin}, {Lee}, \& {Kim}}]{jin2015}
{Jin}, M., {Lee}, J.-E., \& {Kim}, K.-T. 2015, \apjs, 219, 2

\bibitem[{{Kim} {et~al.}(2018){Kim}, {Urquhart}, {Wyrowski}, {Menten}, \&
  {Csengeri}}]{kim2018}
{Kim}, W.~J., {Urquhart}, J.~S., {Wyrowski}, F., {Menten}, K.~M., \&
  {Csengeri}, T. 2018, \aap, 616, A107

\bibitem[{{Kim} {et~al.}(2017){Kim}, {Wyrowski}, {Urquhart}, {Menten}, \&
  {Csengeri}}]{kim2017}
{Kim}, W.-J., {Wyrowski}, F., {Urquhart}, J.~S., {Menten}, K.~M., \&
  {Csengeri}, T. 2017, \aap, 602, A37

\bibitem[{{K{\"o}nig} {et~al.}(2017){K{\"o}nig}, {Urquhart}, {Csengeri},
  {Leurini}, {Wyrowski}, {Giannetti}, {Wienen}, {Pillai}, {Kauffmann},
  {Menten}, \& {Schuller}}]{konig2017}
{K{\"o}nig}, C., {Urquhart}, J.~S., {Csengeri}, T., {et~al.} 2017, \aap, 599,
  A139

\bibitem[{{Lee}(1984)}]{lee1984}
{Lee}, L.~C. 1984, \apj, 282, 172

\bibitem[{{Leung} {et~al.}(1984){Leung}, {Herbst}, \& {Huebner}}]{leung1984}
{Leung}, C.~M., {Herbst}, E., \& {Huebner}, W.~F. 1984, \apjs, 56, 231

\bibitem[{{Lucas}(1976)}]{lucas1976}
{Lucas}, R. 1976, \aap, 46, 473

\bibitem[{{Maluendes} {et~al.}(1993){Maluendes}, {McLean}, {Yamashita}, \&
  {Herbst}}]{maluendes1993}
{Maluendes}, S.~A., {McLean}, A.~D., {Yamashita}, K., \& {Herbst}, E. 1993,
  \jcp, 99, 2812

\bibitem[{{McEwan} {et~al.}(1999){McEwan}, {Scott}, {Adams}, {Babcock},
  {Terzieva}, \& {Herbst}}]{mcewan1999}
{McEwan}, M.~J., {Scott}, G.~B.~I., {Adams}, N.~G., {et~al.} 1999, \apj, 513,
  287

\bibitem[{{Miettinen}(2020)}]{miettinen2020}
{Miettinen}, O. 2020, \aap, 639, A65

\bibitem[{{Mookerjea} {et~al.}(2012){Mookerjea}, {Hassel}, {Gerin}, {Giesen},
  {Stutzki}, {Herbst}, {Black}, {Goldsmith}, {Menten}, {Kre{\l}owski}, {De
  Luca}, {Csengeri}, {Joblin}, {Ka{\'z}mierczak}, {Schmidt}, {Goicoechea}, \&
  {Cernicharo}}]{mookerjea2012}
{Mookerjea}, B., {Hassel}, G.~E., {Gerin}, M., {et~al.} 2012, \aap, 546, A75

\bibitem[{{M{\"u}ller} {et~al.}(2001){M{\"u}ller}, {Thorwirth}, {Roth}, \&
  {Winnewisser}}]{muller2001}
{M{\"u}ller}, H.~S.~P., {Thorwirth}, S., {Roth}, D.~A., \& {Winnewisser}, G.
  2001, \aap, 370, L49

\bibitem[{{Murga} {et~al.}(2020){Murga}, {Kirsanova}, {Vasyunin}, \&
  {Pavlyuchenkov}}]{murga2020}
{Murga}, M.~S., {Kirsanova}, M.~S., {Vasyunin}, A.~I., \& {Pavlyuchenkov},
  Y.~N. 2020, \mnras [\eprint[arXiv]{2007.06568}]

\bibitem[{{Myers} {et~al.}(1996){Myers}, {Mardones}, {Tafalla}, {Williams}, \&
  {Wilner}}]{myers1996}
{Myers}, P.~C., {Mardones}, D., {Tafalla}, M., {Williams}, J.~P., \& {Wilner},
  D.~J. 1996, \apjl, 465, L133+

\bibitem[{{P{\'e}rez-Beaupuits} {et~al.}(2012){P{\'e}rez-Beaupuits},
  {Wiesemeyer}, {Ossenkopf}, {Stutzki}, {G{\"u}sten}, {Simon}, {H{\"u}bers}, \&
  {Ricken}}]{perez-beaupuits2012}
{P{\'e}rez-Beaupuits}, J.~P., {Wiesemeyer}, H., {Ossenkopf}, V., {et~al.} 2012,
  \aap, 542, L13

\bibitem[{{Pety}(2005)}]{pety2005_gildas}
{Pety}, J. 2005, in SF2A-2005: Semaine de l'Astrophysique Francaise, ed.
  F.~{Casoli}, T.~{Contini}, J.~M. {Hameury}, \& L.~{Pagani}, 721

\bibitem[{{Pety} {et~al.}(2005){Pety}, {Teyssier}, {Foss{\'e}}, {Gerin},
  {Roueff}, {Abergel}, {Habart}, \& {Cernicharo}}]{pety2005}
{Pety}, J., {Teyssier}, D., {Foss{\'e}}, D., {et~al.} 2005, \aap, 435, 885

\bibitem[{{Pickett} {et~al.}(1998){Pickett}, {Poynter}, {Cohen}, {Delitsky},
  {Pearson}, \& {M{\"u}ller}}]{pickett1998}
{Pickett}, H.~M., {Poynter}, R.~L., {Cohen}, E.~A., {et~al.} 1998, \jqsrt, 60,
  883

\bibitem[{{Pilleri} {et~al.}(2013){Pilleri}, {Trevi{\~n}o-Morales}, {Fuente},
  {Joblin}, {Cernicharo}, {Gerin}, {Viti}, {Bern{\'e}}, {Goicoechea}, {Pety},
  {Gonzalez-Garc{\'{\i}}a}, {Montillaud}, {Ossenkopf}, {Kramer},
  {Garc{\'{\i}}a-Burillo}, {Le Petit}, \& {Le Bourlot}}]{pilleri2013}
{Pilleri}, P., {Trevi{\~n}o-Morales}, S., {Fuente}, A., {et~al.} 2013, \aap,
  554, A87

\bibitem[{{Rathborne} {et~al.}(2016){Rathborne}, {Whitaker}, {Jackson},
  {Foster}, {Contreras}, {Stephens}, {Guzm{\'a}n}, {Longmore}, {Sanhueza},
  {Schuller}, {Wyrowski}, \& {Urquhart}}]{rathborne2016}
{Rathborne}, J.~M., {Whitaker}, J.~S., {Jackson}, J.~M., {et~al.} 2016, \pasa,
  33, e030

\bibitem[{{Rawlings} {et~al.}(2000){Rawlings}, {Taylor}, \&
  {Williams}}]{rawlings2000}
{Rawlings}, J.~M.~C., {Taylor}, S.~D., \& {Williams}, D.~A. 2000, \mnras, 313,
  461

\bibitem[{{Rizzo} {et~al.}(2005){Rizzo}, {Fuente}, \&
  {Garc{\'{\i}}a-Burillo}}]{rizzo2005}
{Rizzo}, J.~R., {Fuente}, A., \& {Garc{\'{\i}}a-Burillo}, S. 2005, \apj, 634,
  1133

\bibitem[{{Rizzo} {et~al.}(2003){Rizzo}, {Fuente}, {Rodr{\'{\i}}guez-Franco},
  \& {Garc{\'{\i}}a-Burillo}}]{rizzo2003}
{Rizzo}, J.~R., {Fuente}, A., {Rodr{\'{\i}}guez-Franco}, A., \&
  {Garc{\'{\i}}a-Burillo}, S. 2003, \apjl, 597, L153

\bibitem[{{Rodriguez-Franco} {et~al.}(1998){Rodriguez-Franco},
  {Martin-Pintado}, \& {Fuente}}]{rodriguez-franco1998}
{Rodriguez-Franco}, A., {Martin-Pintado}, J., \& {Fuente}, A. 1998, \aap, 329,
  1097

\bibitem[{{Sanhueza} {et~al.}(2012){Sanhueza}, {Jackson}, {Foster}, {Garay},
  {Silva}, \& {Finn}}]{sanhueza2012}
{Sanhueza}, P., {Jackson}, J.~M., {Foster}, J.~B., {et~al.} 2012, \apj, 756, 60

\bibitem[{{Savage} \& {Ziurys}(2004)}]{savage2004}
{Savage}, C. \& {Ziurys}, L.~M. 2004, \apj, 616, 966

\bibitem[{{Schenewerk} {et~al.}(1988){Schenewerk}, {Jewell}, {Snyder},
  {Hollis}, \& {Ziurys}}]{schenewerk1988}
{Schenewerk}, M.~S., {Jewell}, P.~R., {Snyder}, L.~E., {Hollis}, J.~M., \&
  {Ziurys}, L.~M. 1988, \apj, 328, 785

\bibitem[{{Schilke} {et~al.}(2001){Schilke}, {Pineau des For{\^e}ts},
  {Walmsley}, \& {Mart{\'{\i}}n-Pintado}}]{schilke2001}
{Schilke}, P., {Pineau des For{\^e}ts}, G., {Walmsley}, C.~M., \&
  {Mart{\'{\i}}n-Pintado}, J. 2001, \aap, 372, 291

\bibitem[{{Schilke} {et~al.}(1992){Schilke}, {Walmsley}, {Pineau Des Forets},
  {Roueff}, {Flower}, \& {Guilloteau}}]{schilke1992}
{Schilke}, P., {Walmsley}, C.~M., {Pineau Des Forets}, G., {et~al.} 1992, \aap,
  256, 595

\bibitem[{{Sch{\"o}ier} {et~al.}(2005){Sch{\"o}ier}, {van der Tak}, {van
  Dishoeck}, \& {Black}}]{schoier2005}
{Sch{\"o}ier}, F.~L., {van der Tak}, F.~F.~S., {van Dishoeck}, E.~F., \&
  {Black}, J.~H. 2005, \aap, 432, 369

\bibitem[{{Simon} {et~al.}(1997){Simon}, {Stutzki}, {Sternberg}, \&
  {Winnewisser}}]{simon1997}
{Simon}, R., {Stutzki}, J., {Sternberg}, A., \& {Winnewisser}, G. 1997, \aap,
  327, L9

\bibitem[{{Smith} {et~al.}(2012){Smith}, {Shetty}, {Stutz}, \&
  {Klessen}}]{smith2012}
{Smith}, R.~J., {Shetty}, R., {Stutz}, A.~M., \& {Klessen}, R.~S. 2012, \apj,
  750, 64

\bibitem[{{Sternberg} \& {Dalgarno}(1995)}]{sternberg1995}
{Sternberg}, A. \& {Dalgarno}, A. 1995, \apjs, 99, 565

\bibitem[{{Teyssier} {et~al.}(2004){Teyssier}, {Foss{\'e}}, {Gerin}, {Pety},
  {Abergel}, \& {Roueff}}]{teyssier2004}
{Teyssier}, D., {Foss{\'e}}, D., {Gerin}, M., {et~al.} 2004, \aap, 417, 135

\bibitem[{{Tielens}(2013)}]{tielens2013}
{Tielens}, A.~G.~G.~M. 2013, {Interstellar PAHs and Dust}, ed. T.~D. {Oswalt}
  \& G.~{Gilmore}, 499

\bibitem[{{Tielens} \& {Hollenbach}(1985)}]{tielens1985}
{Tielens}, A.~G.~G.~M. \& {Hollenbach}, D. 1985, \apj, 291, 722

\bibitem[{{Tielens} \& {Whittet}(1997)}]{tielens1997}
{Tielens}, A.~G.~G.~M. \& {Whittet}, D.~C.~B. 1997, in IAU Symposium, Vol. 178,
  IAU Symposium, ed. E.~F. {van Dishoeck}, 45

\bibitem[{{Tiwari} {et~al.}(2019){Tiwari}, {Menten}, {Wyrowski},
  {P{\'e}rez-Beaupuits}, {Lee}, \& {Kim}}]{tiwari2019}
{Tiwari}, M., {Menten}, K.~M., {Wyrowski}, F., {et~al.} 2019, \aap, 626, A28

\bibitem[{{Urquhart} {et~al.}(2014){Urquhart}, {Csengeri}, {Wyrowski},
  {Schuller}, {Bontemps}, {Bronfman}, {Menten}, {Walmsley}, {Contreras},
  {Beuther}, {Wienen}, \& {Linz}}]{urquhart2014_atlas_cata}
{Urquhart}, J.~S., {Csengeri}, T., {Wyrowski}, F., {et~al.} 2014, \aap, 568,
  A41

\bibitem[{{Urquhart} {et~al.}(2019){Urquhart}, {Figura}, {Wyrowski},
  {Giannetti}, {Kim}, {Wienen}, {Leurini}, {Pillai}, {Csengeri}, {Gibson},
  {Menten}, {Moore}, \& {Thompson}}]{urquhart2019}
{Urquhart}, J.~S., {Figura}, C., {Wyrowski}, F., {et~al.} 2019, \mnras, 484,
  4444

\bibitem[{{Urquhart} {et~al.}(2018){Urquhart}, {K{\"o}nig}, {Giannetti},
  {Leurini}, {Moore}, {Eden}, {Pillai}, {Thompson}, {Braiding}, {Burton},
  {Csengeri}, {Dempsey}, {Figura}, {Froebrich}, {Menten}, {Schuller}, {Smith},
  \& {Wyrowski}}]{urquhart2018}
{Urquhart}, J.~S., {K{\"o}nig}, C., {Giannetti}, A., {et~al.} 2018, \mnras,
  473, 1059

\bibitem[{{Urquhart} {et~al.}(2013){Urquhart}, {Thompson}, {Moore}, {Purcell},
  {Hoare}, {Schuller}, {Wyrowski}, {Csengeri}, {Menten}, {Lumsden}, {Kurtz},
  {Walmsley}, {Bronfman}, {Morgan}, {Eden}, \& {Russeil}}]{urquhart2013b}
{Urquhart}, J.~S., {Thompson}, M.~A., {Moore}, T.~J.~T., {et~al.} 2013, \mnras,
  435, 400

\bibitem[{{Useli Bacchitta} \& {Joblin}(2007)}]{useli_bacchitta2007}
{Useli Bacchitta}, F. \& {Joblin}, C. 2007, in Molecules in Space and
  Laboratory, 89

\bibitem[{{Vasyunina} {et~al.}(2011){Vasyunina}, {Linz}, {Henning},
  {Zinchenko}, {Beuther}, \& {Voronkov}}]{vasyunina2011}
{Vasyunina}, T., {Linz}, H., {Henning}, T., {et~al.} 2011, \aap, 527, A88

\bibitem[{{Viti} {et~al.}(2004){Viti}, {Collings}, {Dever}, {McCoustra}, \&
  {Williams}}]{viti2004}
{Viti}, S., {Collings}, M.~P., {Dever}, J.~W., {McCoustra}, M.~R.~S., \&
  {Williams}, D.~A. 2004, \mnras, 354, 1141

\bibitem[{{Watt}(1983)}]{watt1983}
{Watt}, G.~D. 1983, \mnras, 205, 321

\bibitem[{{Wienen} {et~al.}(2012){Wienen}, {Wyrowski}, {Schuller}, {Menten},
  {Walmsley}, {Bronfman}, \& {Motte}}]{wienen2012}
{Wienen}, M., {Wyrowski}, F., {Schuller}, F., {et~al.} 2012, \aap, 544, A146

\bibitem[{{Willacy} \& {Williams}(1993)}]{willacy1993}
{Willacy}, K. \& {Williams}, D.~A. 1993, \mnras, 260, 635

\bibitem[{{Wyrowski} {et~al.}(2016){Wyrowski}, {G{\"u}sten}, {Menten},
  {Wiesemeyer}, {Csengeri}, {Heyminck}, {Klein}, {K{\"o}nig}, \&
  {Urquhart}}]{wyrowski2016}
{Wyrowski}, F., {G{\"u}sten}, R., {Menten}, K.~M., {et~al.} 2016, \aap, 585,
  A149

\bibitem[{{Zinnecker} \& {Yorke}(2007)}]{zinnecker2007}
{Zinnecker}, H. \& {Yorke}, H.~W. 2007, \araa, 45, 481

\end{thebibliography}

\onecolumn
\begin{appendix}

\clearpage
\section{Detection rates of molecular lines with source distance}
\begin{figure}[h!]
\centering
\includegraphics[width=0.80\textwidth]{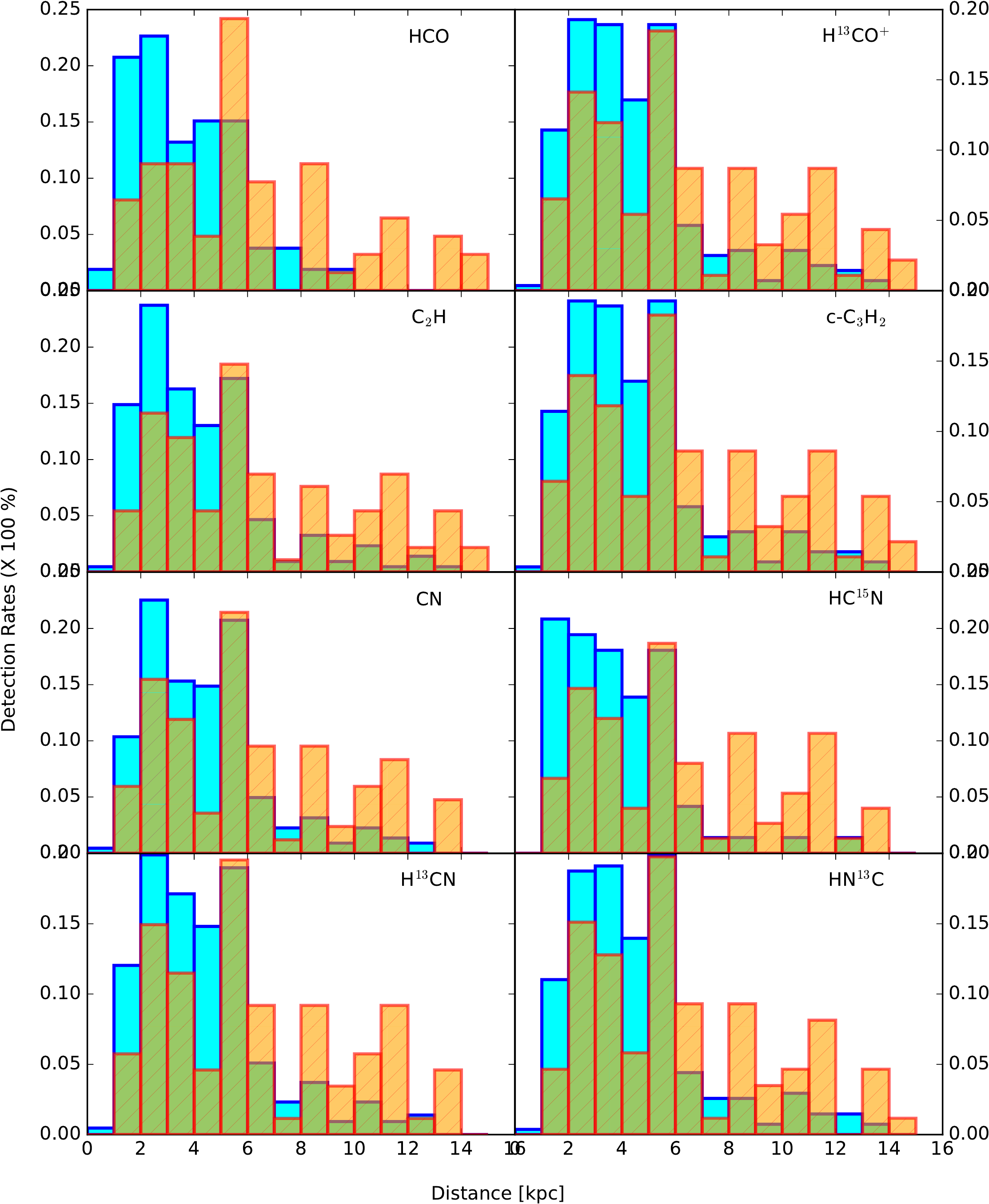}
\caption{\label{appendix:dr_dist} Detection rate as a function of distance.}
\end{figure}
\clearpage
\section{Linewidth of molecular lines with source distance}
\begin{figure}[!ht]
\centering
\includegraphics[width=0.8\textwidth]{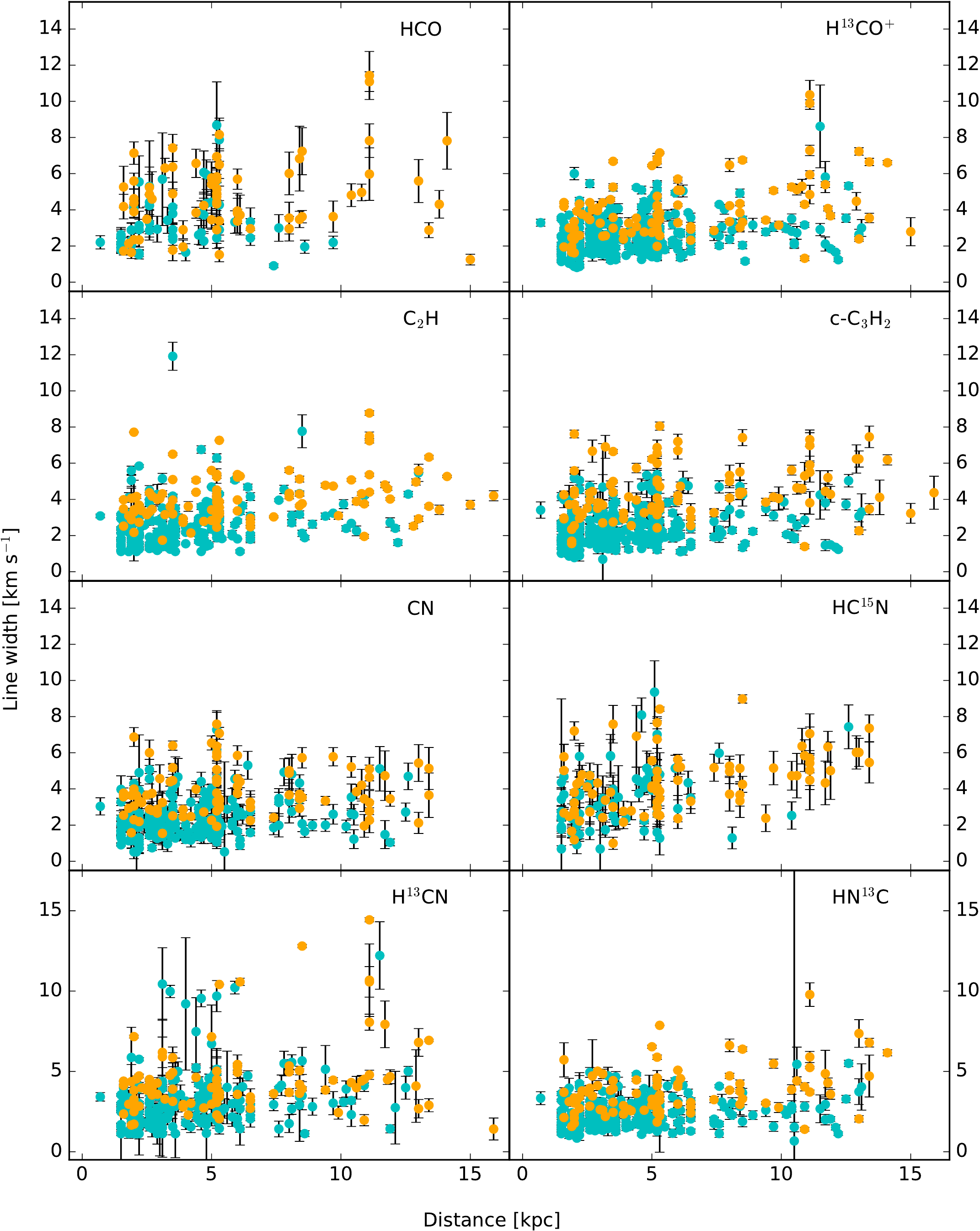}
\caption{\label{appendix:w_dist} Line width as a function of distance.}
\end{figure}

\clearpage
\section{Integrated intensity ratios}
\begin{figure}[!ht]
\centering
\includegraphics[width=0.8\textwidth]{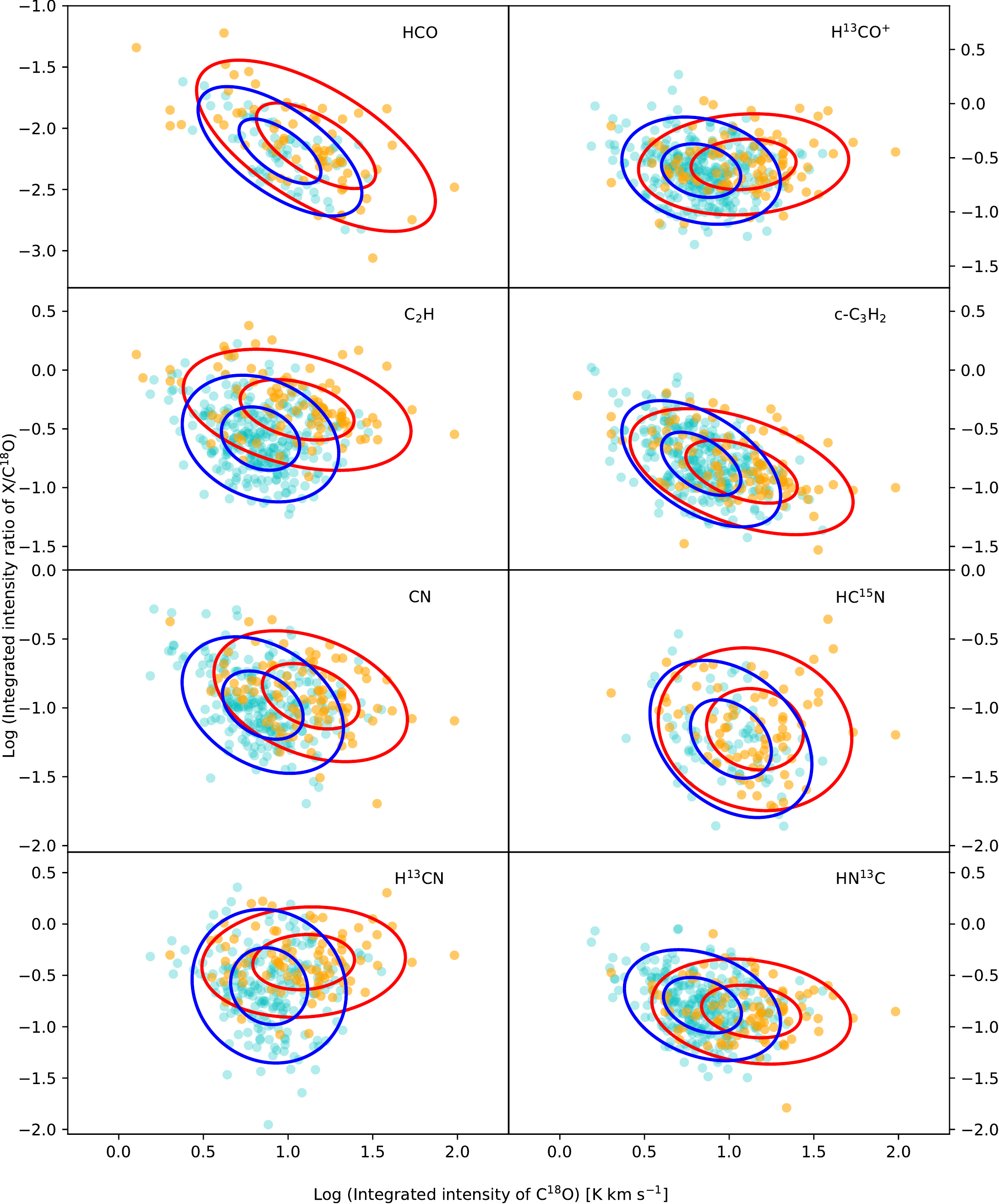}
\caption[]{ \label{appendix:flux_ratio_scatter}  Integrated intensity ratio of X/C$^{18}$O as a function of integrated intensity of C$^{18}$O. The X refers noted molecular species. The orange and cyan symbols indicate \hii\ and non-\hii\ regions, respectively. The red (\hii\ regions) and blue (non-\hii\ regions) eclipses represent covariance error ellipses showing the distributions of the data points.}
\end{figure}

\let\cleardoublepage\clearpage

\end{appendix}

\end{document}